\documentclass[twocolumn]{aastex62}

\received{xx xx xxxx}
\revised{xx xx xxxx}
\accepted{xx xx xxxx}
\submitjournal{ApJ}
\shorttitle{Searching for SMBHBs in the X-rays}
\shortauthors{T. Liu et al.}

\usepackage{epsfig}
\usepackage{xcolor}

\begin{document}

\title{The BAT AGN Spectroscopic Survey -- XVIII. Searching for Supermassive Black Hole Binaries in the X-rays}

\author[0000-0001-5766-4287]{Tingting Liu}
\affiliation{Center for Gravitation, Cosmology and Astrophysics, Department of Physics, University of Wisconsin-Milwaukee, P.O. Box 413, Milwaukee, WI 53201, USA}
\email{tingtliu@uwm.edu}

\author{Michael Koss}
\affiliation{Eureka Scientific, 2452 Delmer Street Suite 100, Oakland, CA 94602-3017, USA}

\author{Laura Blecha}
\affiliation{Department of Physics, University of Florida, Gainesville, FL, USA}

\author{Claudio Ricci}
\affiliation{N\'{u}cleo de Astronom\'{i}a de la Facultad de Ingenier\'{i}a, Universidad Diego Portales, Santiago, Chile}
\affiliation{Kavli Institute for Astronomy and Astrophysics, Peking University, Beijing 100871, China}
\affiliation{George Mason University, Department of Physics \& Astronomy, MS 3F3, 4400 University Drive, Fairfax, VA 22030, USA}

\author{Benny Trakhtenbrot}
\affiliation{School of Physics and Astronomy, Tel Aviv University, Tel Aviv 69978, Israel}

\author{Richard Mushotzky}
\affiliation{Department of Astronomy, University of Maryland, College Park, MD 20742, USA}

\author{Fiona Harrison}
\affiliation{Cahill Center for Astronomy and Astrophysics, California Institute of Technology, Pasadena, CA 91125, USA}

\author{Kohei Ichikawa}
\affiliation{Frontier Research Institute for Interdisciplinary Sciences, Tohoku University, Sendai 980-8578, Japan}

\author{Darshan Kakkad}
\affiliation{European Southern Observatory, Karl-Schwarzschild-Str. 2, 85748, Garching bei München, Germany}

\author{Kyuseok Oh}
\affiliation{Department of Astronomy, Kyoto University, Kyoto 606-8502, Japan}
\affiliation{JSPS Fellow}
\affiliation{Korea Astronomy \& Space Science institute, 776, Daedeokdae-ro, Yuseong-gu, Daejeon 34055, Republic of Korea }

\author{Meredith Powell}
\affiliation{Department of Physics, Yale University, P.O. Box 2018120, New Haven, CT 06520-8120, USA}

\author[0000-0003-3474-1125]{George C. Privon}
\affiliation{Department of Astronomy, University of Florida, 211 Bryant Space Sciences Center, Gainesville, FL 32611, USA}

\author{Kevin Schawinski}
\affiliation{Department of Physics, ETH Zurich, Wolfgang-Pauli-Strasse 27, CH-8093 Zurich, Switzerland}

\author{T. Taro Shimizu}
\affiliation{Max-Planck-Institut f{\"u}r extraterrestrische Physik, Postfach 1312, 85741, Garching, Germany}

\author{Krista Lynne Smith}
\affiliation{KIPAC at SLAC, Stanford University, Menlo Park CA 94025, USA}
\affiliation{Einstein Fellow}

\author{Daniel Stern}
\affiliation{Jet Propulsion Laboratory, California Institute of Technology, 4800 Oak Grove Drive, MS 169-224, Pasadena, CA 91109, USA}

\author{Ezequiel Treister}
\affiliation{Instituto de Astrof\'{i}sica, Facultad de F\'{i}sica, Pontificia Universidad Cat\'{o}lica de Chile, Casilla 306, Santiago 22, Chile}

\author{C. Megan Urry}
\affiliation{Department of Physics, Yale University, P.O. Box 2018120, New Haven, CT 06520-8120, USA}


\begin{abstract}
Theory predicts that a supermassive black hole binary (SMBHB) could be observed as a luminous active galactic nucleus (AGN) that periodically varies on the order of its orbital timescale. In X-rays, periodic variations could be caused by mechanisms including relativistic Doppler boosting and shocks. Here we present the first systematic search for periodic AGNs using $941$ hard X-ray light curves (14--195 keV) from the first 105 months of the \emph{Swift} Burst Alert Telescope (BAT) survey (2004-2013). We do not find evidence for periodic AGNs in \emph{Swift}-BAT, including the previously reported SMBHB candidate MCG+11$-$11$-$032. We find that the null detection is consistent with the combination of the upper-limit binary population in AGNs in our adopted model, their expected periodic variability amplitudes, and the BAT survey characteristics. We have also investigated the detectability of SMBHBs against normal AGN X-ray variability in the context of the {\it eROSITA} survey. Under our assumptions of a binary population and the periodic signals they produce which have long periods of hundreds of days, up to $13$\% true periodic binaries can be robustly distinguished from normal variable AGNs with the ideal uniform sampling. However, we demonstrate that realistic {\it eROSITA} sampling is likely to be insensitive to long-period binaries because longer observing gaps reduce their detectability. In contrast, large observing gaps do not diminish the prospect of detecting binaries of short, few-day periods, as 19\% can be successfully recovered, the vast majority of which can be identified by the first half of the survey.
\end{abstract}



\section{Introduction} \label{sec:intro}

Supermassive black hole binaries (SMBHBs) are expected as the result of galaxy mergers (e.g. \citealt{Begelman1980}), and yet compelling evidence for close-separation SMBHBs has been elusive. Several studies in the past few years have searched for periodically varying quasars as possible SMBHBs in optical time domain surveys, and numerous candidates have been reported \citep{Graham2015Nat,Graham2015,Charisi2016,Liu2015,Liu2016,Liu2019}. These SMBHB candidates typically have (observed) variability periods of $\sim$ a few hundred days -- thousand days. Assuming circular, Keplerian orbits, their masses ($\log(M_{\rm BH}/M_\odot)\sim8-10$) and periods correspond to binary separations of $\sim$ centi- to milli-pc, which are several orders of magnitude more compact than the scale that very long baseline interferometry observations are able to resolve currently (e.g. the radio galaxy 0402+379; \citealt{Rodriguez2006}) or in the future \citep{D'Orazio2018,Burke-Spolaor2018}. These searches for AGN periodicities have been motivated by hydrodynamic and magneto-hydrodynamic simulations which show that accretion onto an SMBHB varies periodically on the order of the binary orbital timescale (e.g. \citealt{MacFadyen2008,Shi2012,Noble2012,D'Orazio2013,Farris2014,Gold2014,Bowen2018,Bowen2019}): the torque exerted by the binary opens up a cavity in the disk in which the binary is embedded (``circumbinary disk''), and the gas instead crosses the gap from the inner edge of the circumbinary disk in the form of narrow streams. The gas eventually feeds the individual accretion disks formed around the black holes (``minidisks'') at a rate that is modulated on the binary orbital timescale. Another possible mechanism is when the steady emission from the minidisk is relativistic Doppler boosted for a highly-inclined, close-separation binary system \citep{D'Orazio2015}, which modulates the apparent flux on the binary orbital period. In both mechanisms, periodicities occur for a wide range of mass ratios ($q$) expected from major mergers: hydrodynamic variability is strongest when $q\gtrsim0.1$, as the binary is able to create an overdensity in the inner edge of the circumbinary disk which interacts with the black holes (e.g. \citealt{Farris2014}). The Doppler boost model favors small mass ratios; however, even in an equal mass binary, periodicity still manifests since the enhancement and the suppression do not cancel.

In the X-ray regime, an SMBHB may also display periodic variability. Similar to the UV/optical, X-ray emissions from the gas bound to the black holes may also experience Doppler boosting \citep{Haiman2017}. Periodicity may also be produced by shocks, when the streams crossing the gap are flung toward by the black holes and hit the cavity wall of the circumbinary disk twice per orbit \citep{Tang2018}. When the stream joins a minidisk, it shock-heats its outer edge and produces bright X-ray emission at tens -- hundreds keV by inverse Compton scattering (\citealt{Roedig2014}, see also \citealt{Farris2015}), which is also modulated on the orbital timescale.

Despite these predictions, the X-ray variability signatures of SMBHBs remain largely unexplored. Only recently was an SMBHB candidate reported by \cite{Severgnini2018} in the center of a Seyfert 2 galaxy at $z=0.0362$, MCG+11$-$11$-$032. Its light curve\footnote{The yet-unpublished data independently analyzed by the Palermo BAT team at \emph{Istituto Nazionale di Astrofisica} (e.g. \citealt{Segreto2010}).} from the Burst Alert Telescope (BAT,  \citealt{Barthelmy2005}) onboard the \emph{Neil Gehrels Swift} Observatory \citep{Gehrels2004} was claimed to show a quasi-periodic variation with a period of about $25$ months over the 123-month baseline, suggesting an orbital velocity of the putative SMBHB: $v \sim 0.06c$. Interestingly, its X-ray spectrum is best described by an absorbed power law with a reflection component plus two narrow Gaussian components for Fe K$\alpha$. While the energy of the second component is less well constrained, their separation in energy $\Delta E$ is consistent with the velocity derived from the variability period, assuming they are produced in either the minidisks or the inner region of the circumbinary disk.

\emph{Swift}-BAT has been observing the hard X-ray sky in the 14--195 keV energy range since its launch. The fifth and the most recent catalog \citep{Oh2018} includes $105$ months of observations from 2004 December to 2013 August and reports $1632$ sources, $947$ of which are un-beamed active galactic nuclei (AGNs). This is currently the largest sample of hard X-ray selected AGNs with long-temporal-baseline, regular observations, and it presents a unique opportunity to study the hard-X ray variability of AGNs.

Previous AGN variability studies with \emph{Swift}-BAT \citep{Beckmann2007,Caballero2012,Shimizu2013,Soldi2014} have measured the fractional flux variability, structure function, or power spectral density (PSD) of BAT AGNs and studied the correlation of variability with physical properties such as black hole mass and X-ray and bolometric luminosities, AGN type, or energy bands. However, no study has so far systematically searched for periodic variability that may be indicative of close SMBHBs at sub-pc separations. Such a search would complement previous ones with ground-based, optical time domain surveys in two important aspects. First, abundant gas is funneled in during (gas-rich) galaxy mergers and thereby powers the SMBHs as luminous quasars (e.g. \citealt{DiMatteo2005Nature,Hopkins2008}), but the SMBHs are likely to be enshrouded by gas and dust, resulting in substantial obscuration in the optical, UV, and even soft X-ray bands (e.g. \citealt{Ricci2017}). Very hard X-ray ($>10$ keV) photons, on the other hand, have a high penetrating power through the obscuring material with column densities upward of 10$^{24}$ cm$^{-2}$ (e.g. \citealt{Ricci2015,Koss2016}) and can therefore potentially reveal SMBH duals and binaries that are inaccessible in the UV/optical (e.g. \citealt{Koss2012,Satyapal2017,Ricci2017,Koss2018}).

Second, while previous numerical simulations of SMBHBs show that the accretion rate from the circumbinary disk onto the minidisks is periodically modulated, it may not directly translate to a periodic photon luminosity due to the buffering effect of the minidisk, when its gas inflow timescale is longer than the modulation timescale (e.g. \citealt{Farris2014}). However, binary-modulated X-ray emission produced by shocks is immune to this effect due to the short timescale of Compton cooling compared to the orbital timescale (see also discussions in e.g. \citealt{Shi2016} and \citealt{Krolik2019}).

The ability of \emph{Swift}-BAT to study the full variable X-ray sky in general and variable AGNs in particular will be unmatched until the extended ROentgen Survey with an Imaging Telescope Array (eROSITA; \citealt{Merloni2012}). It will be much more sensitive to (unobscured) AGNs, many of which will be visited at a high cadence as a result of its scanning strategy.

This paper has the dual goal of performing the first systematic search for periodic AGNs in the X-rays and investigating the prospects for detecting SMBHBs with \emph{eROSITA} and is organized as follows: in Sections \ref{sec:bat}, we search for periodic signals in the 105-month BAT catalog by first modeling the underlying normal AGN variability, which is characterized by higher variability power at lower frequencies (``red noise''). We also revisit the binary candidate MCG+11$-$11$-$032 reported by \cite{Severgnini2018}. In Section \ref{sec:erosita}, we investigate the detectability of periodic SMBHBs with the \emph{eROSITA} survey by first adopting a daily temporal sampling rate over the course of the survey and then proceed to investigate the effects of non-uniform sampling by inserting a gap every 6 months. We further investigate the detectability of short periods with more realistic sampling as a function of the total length of observations. We summarize our results in Section \ref{sec:conclude}.


\section{BAT AGNs}\label{sec:bat}

\subsection{The 105-month \emph{Swift}-BAT Catalog}\label{sec:cat}

\emph{Swift}-BAT has a wide field-of-view (FOV $\sim 60^{\circ}\times100^{\circ}$) and is designed to monitor a comparatively large fraction of the sky for gamma-ray bursts (GRBs). While it is scanning the sky for GRBs and other hard X-ray transients, BAT is also effectively performing a survey of the full sky with a nearly uniform coverage, with $90$\% of the sky covered at the $11$ Ms level over the period of $105$ months, and the median $5\sigma$ sensitivity limit corresponds to $7.24\times10^{-12}$ erg cm$^{-2}$ s$^{-1}$ \citep{Oh2018}. By comparison, the \emph{International Gamma-Ray Astrophysics Laboratory} (INTEGRAL; \citealt{Ubertini2003}) has primarily observed the Galactic plane with its Imager on Board the \emph{INTEGRAL} Satellite (IBIS; \citealt{Winkler2003}) in the $17-100$ keV range with shorter overall exposure time \citep{Bird2010,Krivonos2010}. \emph{NuSTAR} \citep{Harrison2013} has a superior sensitivity in the hard X-ray band (3$\sigma$ sensitivity at $10^{-14}$ erg cm$^{-2}$ s$^{-1}$  in the 10--30 keV range), however it has a smaller field of view (12'.2$\times$12'.2) and mostly performs pointed observations.

Following the earlier survey catalogs \citep{Markwardt2005, Tueller2008, Tueller2010, Baumgartner2013}, the fifth \emph{Swift}-BAT catalog\footnote{https://swift.gsfc.nasa.gov/results/bs105mon/} \citep{Oh2018} contains $1632$ sources observed during the $105$ months between 2004 December and 2013 August, $328$ of which are newly-identified sources since its last version, the $70$-month catalog \citep{Baumgartner2013}. After making a blind source detection at the $4.8\sigma$ level and fitting for the source position, a cross-search using a fixed radius is made in the archive for counterparts observed by other telescopes/instruments such as \emph{Swift}-XRT, \emph{Chandra}, and \emph{XMM-Newton}. The X-ray sources are also searched for optical counterparts in the NED and SIMBAD databases. Sources with known types in their optical counterparts are further divided into classes, and the $105$-month catalog includes $947$ un-beamed AGNs, i.e. classified as Seyfert 1, Seyfert 2, or LINER based on their emission lines as well as ``Unknown AGNs'' . A detailed description of the BAT hard X-ray survey data can be found in \cite{Oh2018} and references therein. One of the main data products is the light curves of BAT-detected sources spanning the duration of the survey. Instead of ``snapshot'' light curves from individual observations, the 105-month catalog presents the monthly-binned light curves, by adding individual snapshot images from each month of the survey and measuring the source flux from the total-band mosaic image.

Follow-up observations and studies of BAT-detected sources are actively being carried out. Among them is the BAT AGN Spectroscopic Survey\footnote{http://www.bass-survey.com} (BASS; \citealt{Koss2017}), a large effort to measure optical spectra for this hard X-ray selected, uniquely-unbiased sample of AGNs with complete estimates for black hole mass, accretion rate, and bolometric luminosity using dedicated spectroscopic observations.  In addition to optical spectra, BASS also presents careful determination of the X-ray properties of BAT AGNs by combining \emph{Swift}-BAT data with observations from a variety of soft X-ray telescopes \citep{Ricci2017BASS}. The BASS sample consists of the brightest (L$_{\rm 2-10 keV}\gtrsim$10$^{42}$ erg s$^{-1}$) and the nearest ($90$\% are at $z<0.2$) AGNs, allowing detailed studies of nearby AGNs while serving as a benchmark for X-ray surveys of a large sample of high-redshift AGNs, such as the upcoming \emph{eROSITA} and the planned spectroscopic follow-up of \emph{eROSITA}-detected AGNs with SDSS-V \citep{Kollmeier2017} and 4MOST \citep{Merloni2019}.


\subsection{Variability Analysis}\label{sec:analysis}

\begin{figure}
\centering
\epsfig{file=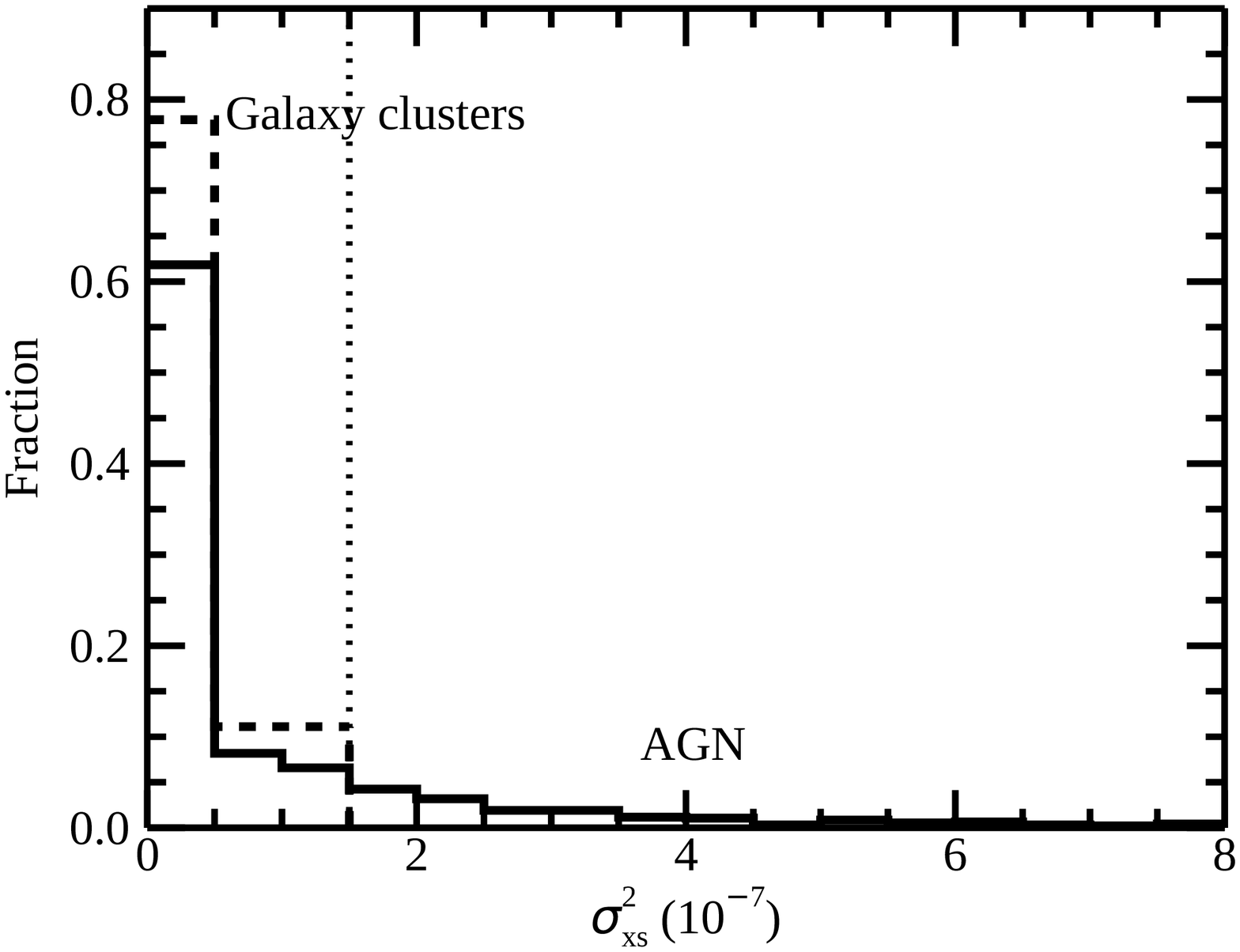,width=0.45\textwidth,clip=}
\epsfig{file=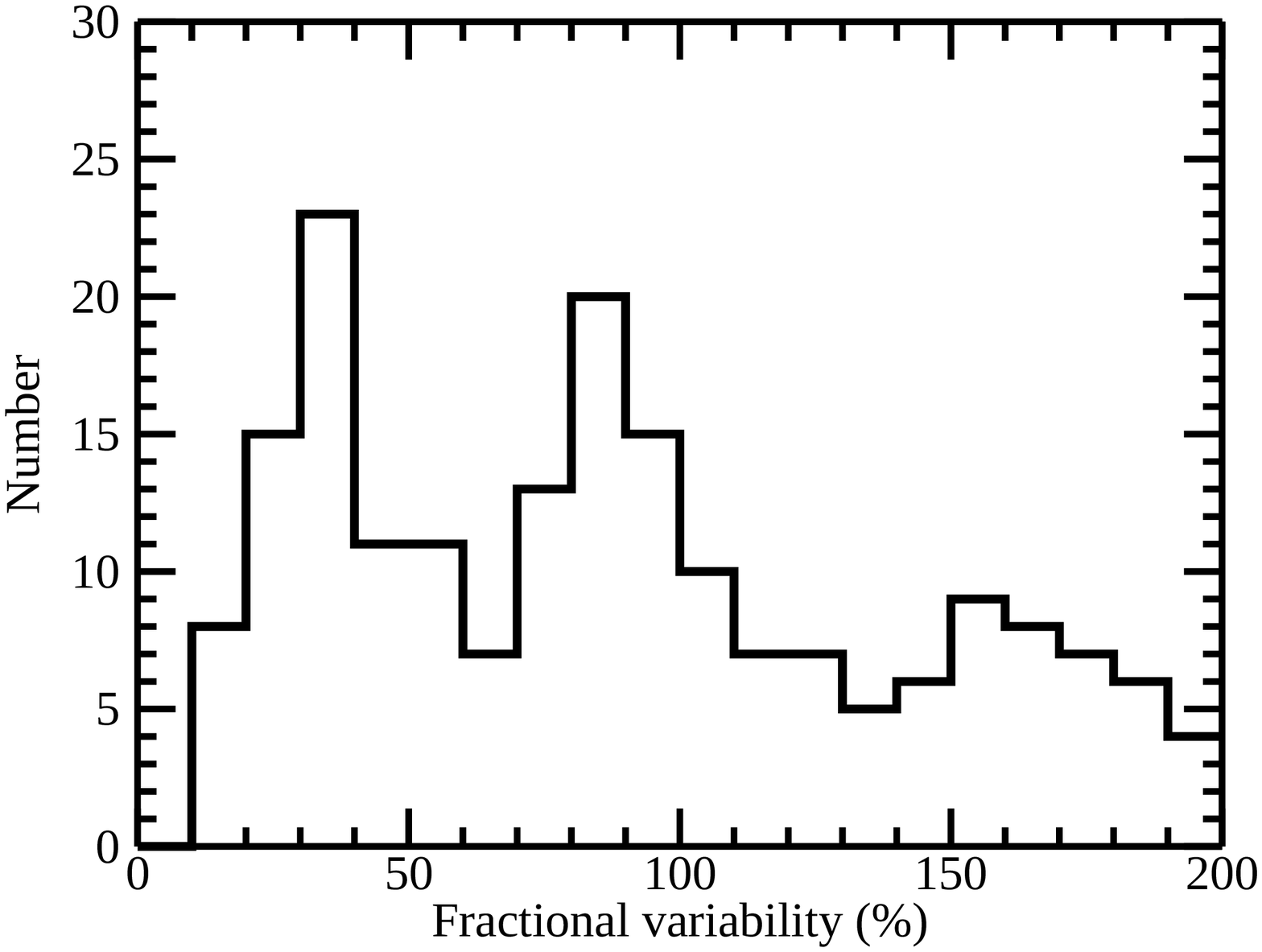,width=0.45\textwidth,clip=}
\caption{Upper panel: the excess variance of the galaxy cluster sample (dashed histogram) versus the full AGN sample (solid histogram). Those with negative excess variance are defined as $\sigma_{\rm xs}^2 = 0$. The variability threshold is marked with a dotted line. Lower panel: the fractional variability distribution of the variable AGN sample. AGNs with fractional variability larger than $200$\% are defined to have $f_{\rm var} = 200$\% (see text).}
\label{fig:sigma_xs}
\end{figure}

Our parent sample consists of $941$ un-beamed AGNs in the $105$-month catalog that are classified as `Seyfert 1', `Seyfert 2' or `Unknown AGN'. Their source IDs, names, and coordinates are listed in Table \ref{tab:sample}. We first calculate the excess variance \citep{Nandra1997,Edelson2002,Vaughan2003} for each light curve, which removes the apparent variation due to measurement errors: $\sigma^{2}_{\rm xs}$ = S$^{2}$ - $\overline{\sigma^{2}_{\rm err}}$, where S$^2$ is the variance of the light curve, and $\sigma_{\rm err}$ is the measurement error. To select intrinsically variable AGNs\footnote{We assume that all AGNs are likely intrinsically variable at some level. Here we will exclude any AGN whose observed variability is largely due to measurement uncertainties.}, we compare their $\sigma^{2}_{\rm xs}$ with those of galaxy clusters, which are constant hard X-ray sources (e.g. \citealt{Wik2011}). We have inspected the \emph{Swift}-XRT and \emph{XMM-Newton} data of each cluster, in order to exclude those with contamination from variable AGNs in the BAT FOV. We have also used the detailed analysis of BAT clusters by \cite{Ajello2009} and \cite{Ajello2010}. This process removed 8/26 clusters. As we show in Figure \ref{fig:sigma_xs} (upper panel), the distribution of AGNs closely mimics that of clusters at the $\sigma^{2}_{\rm xs} < 1.5\times10^{-7}$ level, indicating possible remaining systematic effects; above this level, the fraction of galaxy clusters declines to zero. Therefore, we use $\sigma^{2}_{\rm xs} = 1.5\times10^{-7} $ as the variability threshold and select $220$ AGNs (23\% of the sample) for further analysis. We show the $\sigma^{2}_{\rm xs}$ values of the full sample in Table \ref{tab:sample} and denote those that meet our variability threshold.

\begin{table*}[ht]
\caption{Variability Properties of the BAT AGN Sample}
\begin{center}
\begin{tabular}{llrrrccc}
\hline \hline
ID\tablenotemark{a} & Name\tablenotemark{b} & R.A.\tablenotemark{c} & Decl.\tablenotemark{d} & $\sigma^{2}_{\rm xs}$ ($10^{-7}$)\tablenotemark{e} & Variable?\tablenotemark{f} & $f_{\rm var} (\%)$\tablenotemark{g} & Type\tablenotemark{h} \\
    1  & SWIFT J0001.0-0708   &    0.228  &    -7.164  &   10.6620  &  Y  &  202.09 &    red \\
    2 &  SWIFT J0001.6-7701   &    0.445 &    -77.000  &    0.0296  &  N  &        -   &    - \\
    3 &  SWIFT J0002.5+0323  &     0.613  &     3.365  &  0.0000    &  N    &      -  &    - \\
    4  & SWIFT J0003.3+2737  &     0.856  &    27.643  &    1.2593  &  N   &       -  &     - \\
    5 &  SWIFT J0005.0+7021  &     0.934  &    70.358  &   0.5252   &  N   &       -  &     - \\
    6 &  SWIFT J0006.2+2012   &    1.596 &     20.242  &   0.0000   &  N   &       -  &     - \\
    7 &  SWIFT J0009.4-0037   &    2.305 &     -0.639  &  0.0000    &  N    &      -    &   - \\
   10 &  SWIFT J0021.2-1909  &     5.289 &    -19.162   &    0.0000 &  N  &        -   &    - \\
   13 &  SWIFT J0025.8+6818  &     6.432  &    68.403  &   0.2152   &  N    &      -   &    - \\
   14 &  SWIFT J0026.5-5308   &    6.709 &    -53.151  &   2.0127   &  Y   &  88.70 &    red  \\
\hline \hline
\end{tabular}
\end{center}
\tablecomments{Table 1 is published in its entirety in the machine-readable format. A portion is shown here for guidance regarding its form and content.}
\tablenotetext{a}{Swift-BAT 105-month catalog ID}
\tablenotetext{b}{BAT name of the source}
\tablenotetext{c}{BAT right ascension of the source}
\tablenotetext{d}{BAT declination of the source}
\tablenotetext{e}{Excess variance $\sigma^{2}_{\rm xs}$ = S$^{2}$ - $\overline{\sigma^{2}_{\rm err}}$. (A negative value is forced to be zero.)}
\tablenotetext{f}{Whether the source is classified as intrinsically variable}
\tablenotetext{g}{Fractional variability $f_{\rm var} = (\sigma_{\rm XS}/\small \langle F \small\rangle)\times100$\%}
\tablenotetext{h}{Whether the intrinsic variability can be characterized by red noise or white noise}
\label{tab:sample}
\end{table*}

We have also calculated the fractional variability, which is normalized to the average flux of the source: $f_{\rm var} = (\sigma_{\rm XS}/\big \langle F \big\rangle)\times100$\% (Table \ref{tab:sample}). It is similar to the $S_{\rm V}$ parameter used by \cite{Soldi2014}, where their $\sigma_{\rm Q}$ parameter reduces to $\sigma_{\rm XS}$ for uniform measurement errors. As we show in Figure \ref{fig:sigma_xs} (lower panel), most AGNs in this sample are variable at the $30-40$\% level on the $\sim$month timescale, similar to the sample from \cite{Soldi2014}, although with a heavy tail for AGNs with fractional variability $\gtrsim100$\%. We find that those high-variance values ($f_{\rm var}>50$\%) tend to be associated with low count rates. Additionally, visual inspection of their light curves reveals a number of outliers, and thus we only show $f_{\rm var}\le200$\% in Figure \ref{fig:sigma_xs} for presentation purposes. We confirm that by using $\sigma^2_{\rm xs}$ instead of $f_{\rm var}$ as the variability threshold, we do not systematically bias against faint sources, as the mode count rates of the selected variable AGNs and those of ``non-variable'' AGNs are both $\approx5\times10^{-4}$ cts s$^{-1}$\

Since the 105-month light curves are uniformly sampled on monthly timescales, we compute the periodogram, which is defined as the modulus squared of the Fourier transform and is normalized to have units of (rms/mean)$^{2}$ Hz$^{-1}$. We ignore the point at the Nyquist frequency\footnote{Because the periodogram at this frequency is $\chi_{1}^{2}$-distributed (e.g. \citealt{Vaughan2005}).} and fit a simple linear function to the periodogram in log-space and estimate the power law continuum of the power spectrum, where here we have corrected for the bias between the periodogram and the power spectrum by adding $0.25068$ to the best fit linear function (see \citealt{Papadakis1993,Vaughan2005}). 

\begin{figure}
\centering
\epsfig{file=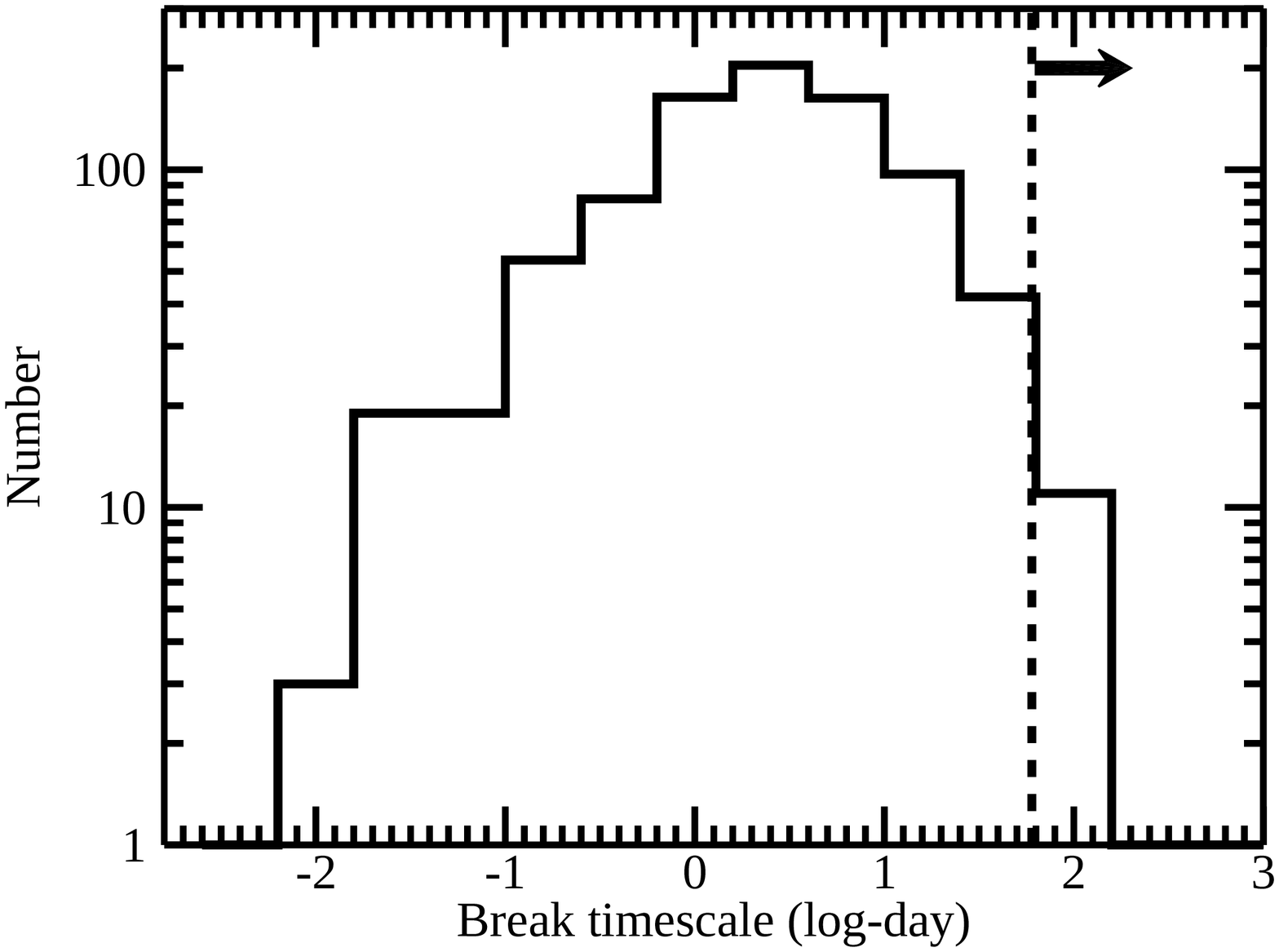,width=0.5\textwidth,clip=}
\caption{The expected PSD break timescales in units of log-day derived from the relation in \cite{Gonzalez-Martin2012} for the BAT AGNs with black hole mass measurements from BASS DR2. Note that the y-axis is log-scale for clarity. The majority (98\%, or $845/861$ AGNs) of break timescales are shorter than the variability timescales probed by BAT ($>2$ months, to the right of the dashed line).}
\label{fig:mbh_tbr}
\end{figure} 

We confirm the goodness-of-fit using the Kuiper's test, since the periodogram should scatter around the true PSD following a $\chi^{2}$ distribution with two degrees of freedom (e.g. \citealt{Vaughan2005}). By the same token, a possible feature with a power below $-\ln[1-(1-\epsilon)^{1/n}]P(f)$ can be rejected as a spurious peak at the ($1-\epsilon$) level, where $\epsilon$ is the chosen false alarm probability, and the trial factor $n$ is the number of frequency points in the range where the periodogram is fitted.

By fitting a simple linear function to the periodogram, we have assumed that the underlying red noise power spectrum is described by a single power law in the frequency range of interest, and here we show that it is indeed a reasonable assumption. Since the break timescale of the X-ray power spectrum is found to correlate with black hole mass \citep{McHardy2006,Gonzalez-Martin2012}, we can use the best-fit correlation in \cite{Gonzalez-Martin2012} to estimate the expected break timescale\footnote{Here we have assumed there is no significant difference in the PSDs in the soft ($<10$ keV) and hard ($>10$ keV) X-ray bands (e.g. \citealt{Shimizu2013}).}: $\log(T_{\rm br}) = 1.09 \log(M_{\rm BH})-1.70$, where $T_{\rm br}$ is in units of days and $M_{\rm BH}$ is in units of 10$^{6}$$M_{\odot}$. We use black hole masses from the internally-released BASS DR2 (which is soon to be publicly available), which has a higher completion percentage for black hole mass measurements ($\approx$90\%) than the published DR1 \citep{Koss2017}. Assuming this sample is representative of all BAT AGNs, we expect the majority ($>98$\%) to have break timescales T$_{\rm br}<2$ month, which corresponds to frequencies that are higher than the Nyquist frequency in our BAT data (Figure \ref{fig:mbh_tbr}). This is also consistent with the lack of detection of PSD breaks and the lower limit at 10$^{-6}$ Hz reported by \cite{Shimizu2013}.

\cite{Shimizu2013} also showed that the brightest AGNs do not become white noise-dominated until the $\sim5$-day timescale, and their power spectra can be well described by single power laws over the full frequency range. However, the fainter sources in our sample show evidence of white noise beginning to dominate at a much lower frequency, and thus we should only fit for the power law continuum in the frequency range where red noise dominates in order to estimate its slope. Hence, for each of the 220 AGNs in our variable sample, we will also consider a ``power law+constant'' model, in addition to a single power law fit (Section \ref{sec:bat105psd}). Ideally, the constant power level should be consistent with the level of Poisson noise estimated from measurement errors: $P_{\rm N} = \frac{2\Delta T\overline{\sigma^{2}_{\rm err}}}{\mu^2}$ \citep{Vaughan2003}, where $\mu$ is the mean count rate and $\Delta T = 1$ month for the 105-month BAT light curve. (For example, the power spectrum of a constant X-ray source such as a galaxy cluster should be approximately flat and at a level that is consistent with $P_{\rm N}$.)


\begin{figure}
\centering
\epsfig{file=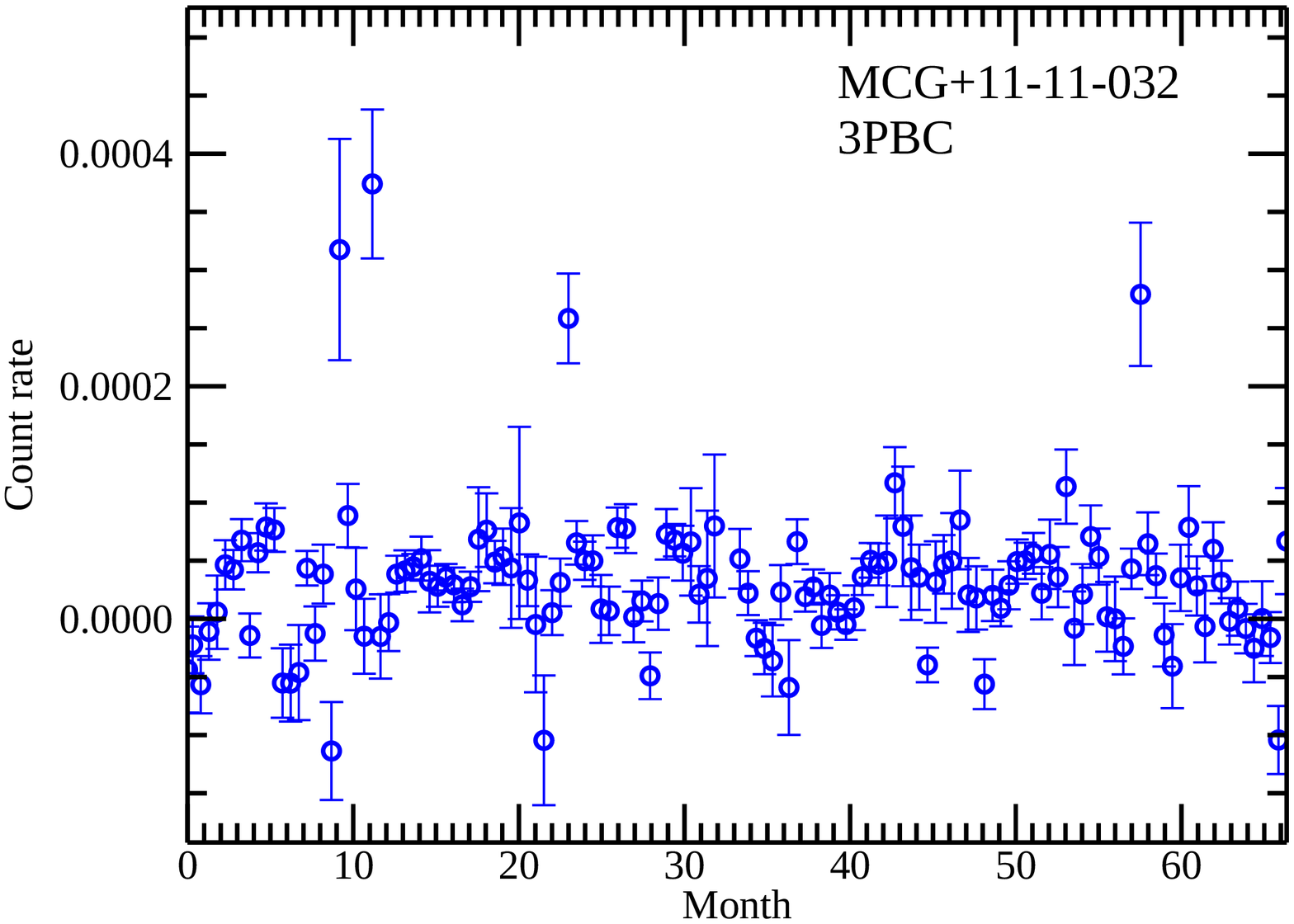,width=0.45\textwidth,clip=}
\epsfig{file=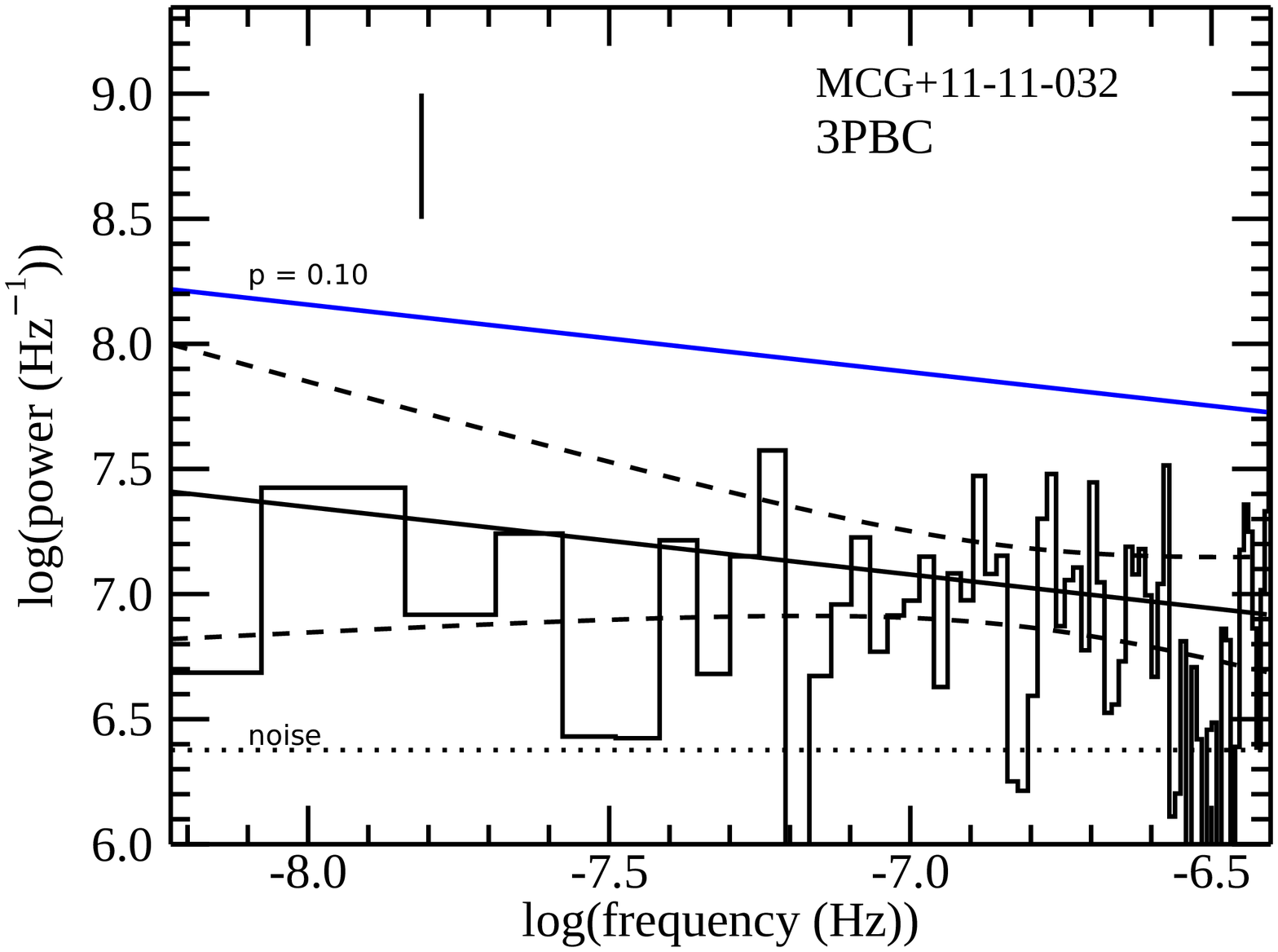,width=0.45\textwidth,clip=}
\caption{Upper panel: the BAT light curve of MCG+11$-$11$-$032 since beginning of the mission (Month = 0). The Third Palermo BAT Catalog (3PBC) publishes light curves from the first 66 months of the survey. Lower panel: its power spectrum. Solid line represents the best fit power law continuum, and dashed lines represent model uncertainties, which are determined from the propagation of the uncertainties of the power law slope, normalization, and their covariance (see \citealt{Vaughan2005}). The estimated noise level is marked with a dotted line. There is no peak at $25$ months (black tick mark) even at the modest $90$\% level (blue solid line). The four outlying points in the upper panel were not removed in our analysis.}
\label{fig:pbc}
\end{figure} 

\subsection{MCG+11$-$11$-$032 Revisited}\label{sec:severgnini}

We first demonstrate this procedure by testing for the reported periodicity in MCG+11$-$11$-$032. Since its light curve presented by \cite{Severgnini2018} was independently analyzed by the Palermo team, we retrieved the light curve from their published Third Palermo BAT Catalog (3PBC) instead of the 105-month catalog (Figure \ref{fig:pbc}). We consider the source red-noise dominated, since its power spectrum is significantly above the estimated noise level and the power law slope is not flat after taking into account its uncertainty. Thus, we measure its power spectrum in the full frequency range and are able to reject signals in the full range at the $>90$\% level, including the putative period of $25$ months ($\log[f/Hz] = -7.8$) reported by \cite{Severgnini2018} (Figure \ref{fig:pbc}). However, we note several differences that may result in our different conclusions: \cite{Severgnini2018} adopted a different method, where the periodic function is only superimposed for visual purposes, and no systematic search or power spectral analysis was performed. Second, their data were independently processed and were not available to us, so our comparison is not a direct one. Finally, their light curve was from the first $123$ months of the survey (which has not been published), and thus robustly detecting two cycles of the signal in the $66$-month 3PBC light curve is more difficult. 


\subsection{The BAT 105-month sample: Power spectrum fitting}\label{sec:bat105psd}

We then apply the method to the full sample of $220$ variable AGNs in the 105-month dataset. We first naively fit a single power law in the full frequency range and obtain the best-fit power law slope, normalization, and their uncertainties. If the slope is steeper than its uncertainty, then we tentatively classify the power spectrum as ``red'', and ``white'' otherwise. We find that all $220$ AGNs can be reasonably described by either a white power spectrum (124/220), or a red single power law power spectrum (96/220 ; see Table \ref{tab:sample}).

However, for the fainter sources, the level of white noise becomes comparable to, or even dominates over, the power law red noise at high frequencies. Hence, in cases where the estimated $P_{\rm N}$ is comparable to the computed power spectrum at high frequencies, we model the power spectrum by fitting it to a power law+constant: we use the naive single power law best-fit normalization parameter and the estimated $P_{\rm N}$ level as respective initial guesses and vary by a step size of $0.1$ (in logarithmic space); we vary the power law slope in the range of [0, 3] in a step size of 0.1.

We note that since we are effectively ``de-reddening'' the power spectrum by first fitting it to a power law and do not aim to measure the ``true'' PSD slope, we refrain from directly comparing our distribution of the best-fit power law slopes with those measured by \cite{Shimizu2013}, except for noting that, qualitatively, our distribution of slopes would contain more flatter slopes than the \cite{Shimizu2013} distribution, due to the high level of white noise of the fainter sources in our sample.


\subsection{The BAT 105-month sample: Upper limits on periodic signals}\label{sec:bat105periodic}

After obtaining reasonable fits to the power spectra, we proceed to apply the method in Section \ref{sec:analysis} to test for periodic signals. We have chosen $99.7\%$ as the significance level, since we do not expect any source in our sample to have peaks above this threshold (i.e. as a false positive). We hence reject the presence of any periodic signal at this level in this sample.

\begin{figure}
\centering
\epsfig{file=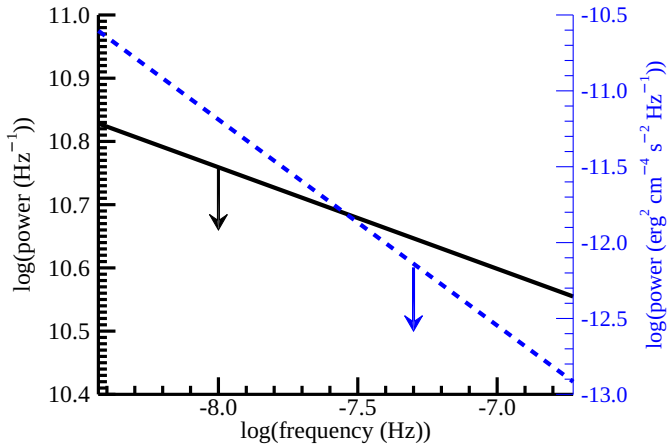,width=0.5\textwidth,clip=}
\caption{The 99.7\% upper limit on a periodic signal determined from the two best sources in the BAT AGN sample, represented here in fractional rms and flux units (black solid and blue dashed lines, respectively). The significance levels are computed based on the best power spectrum fit (see text for details). }
\label{fig:lim}
\end{figure}

The null-detection nevertheless allows us to put upper-limit constraints on periodic amplitudes in the BAT volume as a function of frequency. As we show in Figure \ref{fig:lim}, the most stringent upper limit in log-power units is given by NGC 7214 (represented by the solid line). While the periodogram is conventionally normalized to have fractional rms units, we can also calculate the variability power in terms of ``absolute'' units and convert to an upper limit in physical units, as the BAT count rate is normalized to the Crab ($f_{14-195 \rm keV} = 2.33\times10^{-8}$ erg cm$^{-2}$ s$^{-1}$). Hence, the best upper limit in physical flux units is provided by NGC 2110, which is also one of the brightest AGNs in the sample (dashed line in Figure \ref{fig:lim}). We note that those upper limits apply to a strictly periodic signal, rather than a ``quasi-periodic'' signal, which has a finite width in Fourier frequency.

Given the upper limit of $\sim1$ periodic source per $10^{4}$ AGNs in optical surveys out to a higher redshift \citep{Liu2019}, a null detection in $\sim1000$ BAT AGNs at lower redshifts was to be expected. As we will also show in Section \ref{sec:mock}, the null detection in BAT is consistent with the small amplitudes of binary-induced periodic variability and the large measurement uncertainties of BAT. 

Finally, we note that another previously reported SMBHB candidate in the BAT sample, PKS 1302-102 (or PG 1302-102; hereafter PG 1302), was not recovered in our periodicity search. PG 1302 was proposed as a binary candidate for its smooth, sinusoid-like variation in the optical light curve with a $\sim 5$ yr observed period over $\sim 2$ cycles \citep{Graham2015Nat}, which was attributed to the relativistic Doppler boost from an un-equal mass binary (\citealt{D'Orazio2015}; however see \citealt{Vaughan2016} and \citealt{Liu2018}; see also \citealt{Duffell2019} for an alternative interpretation). Since PG 1302, which is classified as a ``beamed AGN'' in the BAT 105-month catalog, was not included our parent sample, we have performed our light curve analysis separately on this source. We find that its power spectrum shows no evidence for peaks or features and is well-characterized by white noise which is consistent with the expected Poisson noise level.


\section{Detection Prospects for eROSITA}\label{sec:erosita}

While we do not find periodicities in the BAT sample, the upcoming \emph{eROSITA} mission is likely to transform the search for SMBHBs in the X-rays, thanks to its high sensitivity in the 0.5--10 keV band and its all-sky scanning strategy. In this section, we will attempt to investigate the detectability of SMBHBs as periodic AGNs with \emph{eROSITA}.

More specifically, we will test for periodic signals produced by a mock population of binaries that are superimposed on red noise. Assuming a fixed variance for the red noise PSD (see Section \ref{sec:analysis} and discussions below), only two parameters are needed to generate a light curve --- the amplitude and period of the signal. These two elements will then be determined by binary parameters from the mock population. While the eROSITA sky is divided by half between the German and Russian consortia, we refer to a full-sky SMBHB population whenever applies.


\subsection{A Mock SMBHB Sample}\label{sec:mock}

\emph{eROSITA} will be located at the L2 Lagrangian point and scan the sky in great circles, completing one circle in four hours. As its survey plane progresses around the sun by $\sim 1$ deg per day, \emph{eROSITA} will complete a scan of the full sky every $6$ months and eight full-sky scans during its survey lifetime (eRASS1 -- eRASS8). As a result of this scanning strategy, the ecliptic poles are more frequently visited than lower latitudes (more details can be found in \citealt{Merloni2012}). Its observing cadence in this region of the sky, combined with the survey sensitivity, could probe a wide range of variability timescales and thus possible binary parameters.

To investigate this prospect, we first construct a full-sky binary population. We adopt the mock \emph{eROSITA} AGN catalog of \cite{Comparat2019}. The method populates dark matter halos with galaxy stellar masses and then AGNs using an abundance matching technique and reproduces the observed AGN X-ray luminosity function. For the remainder of the section, we adopt the full mock catalog (``eRASS8''), which includes 2.6 million AGNs at $0<z<6$. To convert the galaxy stellar mass to the black hole mass, we use the $M_{\rm stellar}-M_{\rm BH}$ relation from \cite{Reines2015}\footnote{Here we have not considered the possible redshift evolution of the $M_{\rm stellar}-M_{\rm BH}$ relation.}: $\log(M_{\rm BH}/M_{\odot}) = 7.45+1.05 \log(M_{\rm stellar}/10^{11}M_{\odot})$, and the resulting range of $M_{\rm BH}$ is $\sim 10^{6}-10^{8} M_{\odot}$ (we will further discuss this scaling relation and our black hole masses below).

For this study, we only focus on those AGNs near the ecliptic poles which are better sampled than those at lower latitudes. We choose areas that are no more than $2$ degrees away from the poles and select AGNs from the \cite{Comparat2019} catalog that are in those regions using a grid: $\Delta$RA = 0.05 h and $\Delta$dec = 0.5 deg. Our parent sample contains $\sim 6\times10^{3}$ AGNs.

To compute an upper limit on the number of SMBHBs that could exist in this AGN sample, we assume a one-to-one correspondence between an AGN and an SMBHB. This is motivated by the match between the AGN lifetime of $t_{\rm AGN} \sim 10^{7}$ yr and the timescale for a binary to evolve from the outer edge of the circumbinary disk to coalescence \citep{Haiman2009}. We further assume: (1) all binaries in the footprint have evolved into the gravitational wave-emitting regime where the time (``residence time'') a binary spends at an orbit that corresponds to the orbital timescale $t_{\rm orb}$ is $t_{\rm res}$ = 1.11$\times10^{7}$ yr $q_{s}^{-1}M_{7}^{-5/3}t_{\rm orb}^{8/3}$ \citep{Haiman2009}, where $q_{s} = 4q/(1+q)^2$ with $q=M_{2}/M_{1}$ being the mass ratio, $M_{7} = M_{\rm BH}/10^{7}M_{\odot}$, and $t_{\rm orb}$ is in units of yr; (2) all residence timescales can be probed by either Case A or Case B, i.e. $t_{\rm res, min} \sim 10^{5}$yr and $t_{\rm res, max} \sim  10^{7}$yr, and each binary is assigned a residence time according to the linear dependence of the binary number rate on $t_{\rm res}$: $f (t_{\rm res}) = t_{\rm res}/t_{\rm AGN}$ \citep{Haiman2009}. The upper bound of $10^{7}$yr is motivated by the binary evolution timescale as we previously discussed, above which the binaries can no longer be active as AGNs during their entire lifetimes, and our assumptions are no longer valid. The lower bound is such that the expectation number of binaries with the shortest $t_{\rm res}$ is at least a few in a sample of $10^{3}-10^{4}$ AGNs. While in principle, the absolute lower limit on $t_{\rm res}$ is where the separation $a = r_{\rm ISCO}$, we note that our results should be insensitive to the lower bound on $t_{\rm res}$, since binaries with very short residence timescales are exceedingly rare.

\begin{figure*}
\centering
\epsfig{file=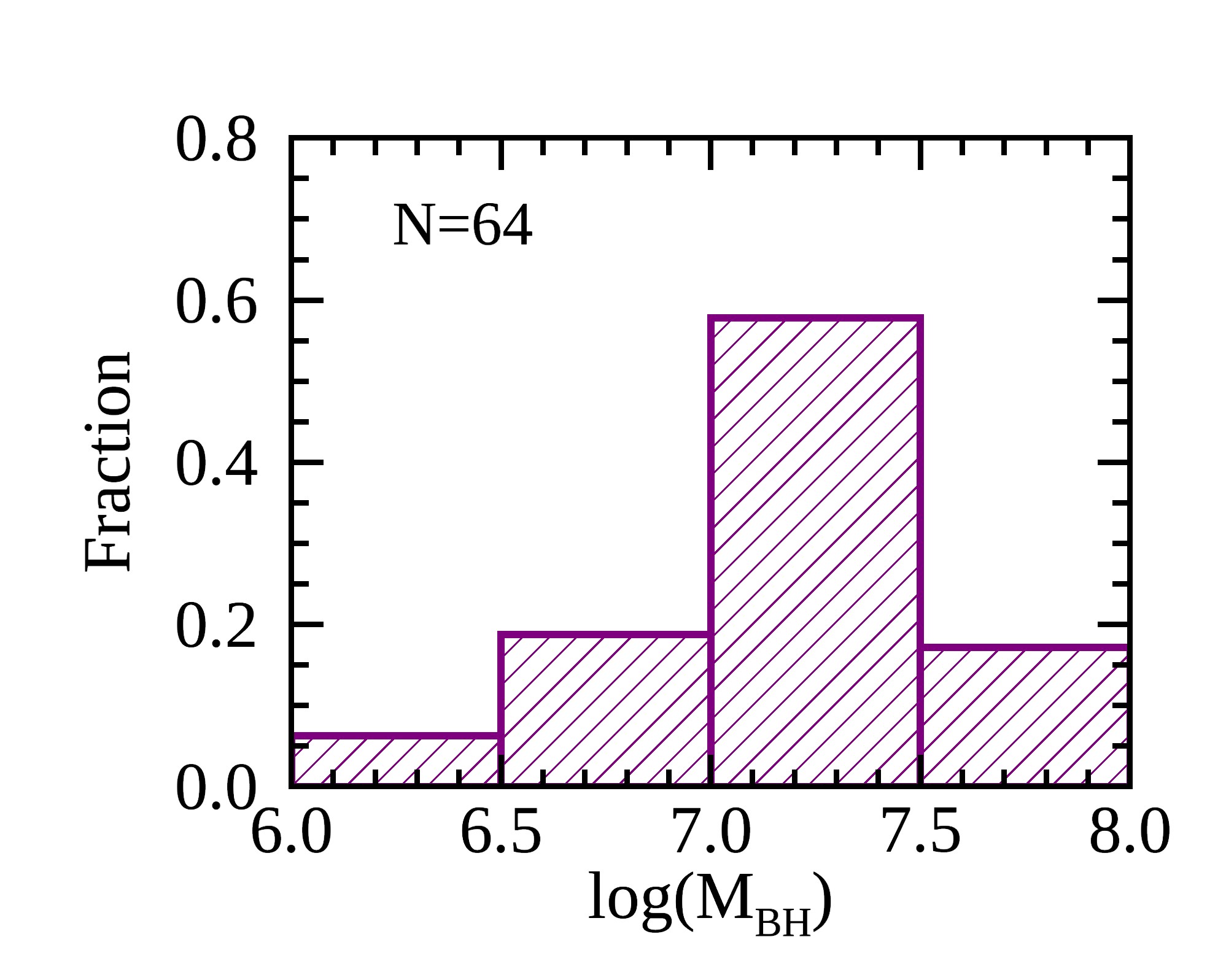,width=0.32\textwidth,clip=}
\epsfig{file=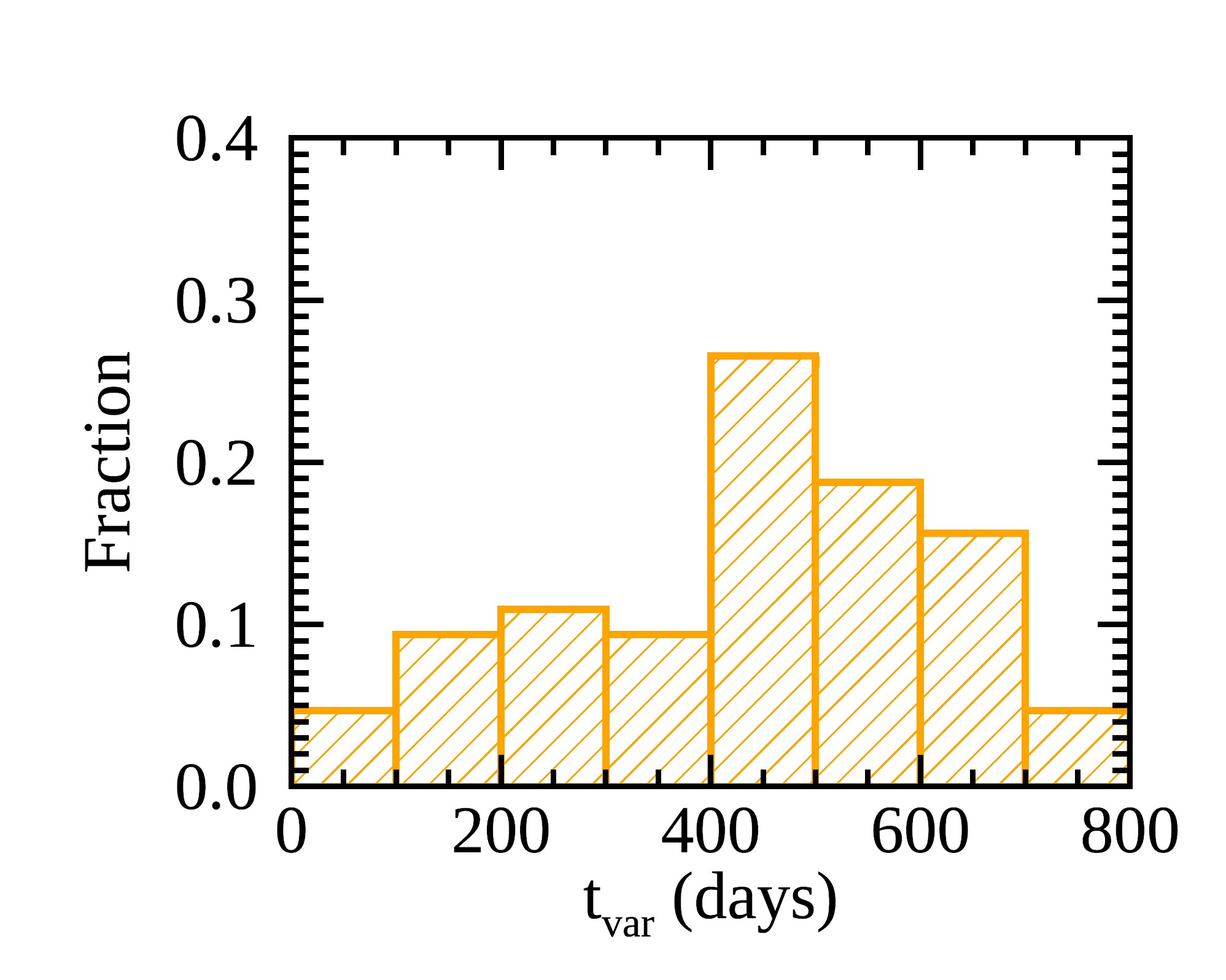,width=0.32\textwidth,clip=}
\epsfig{file=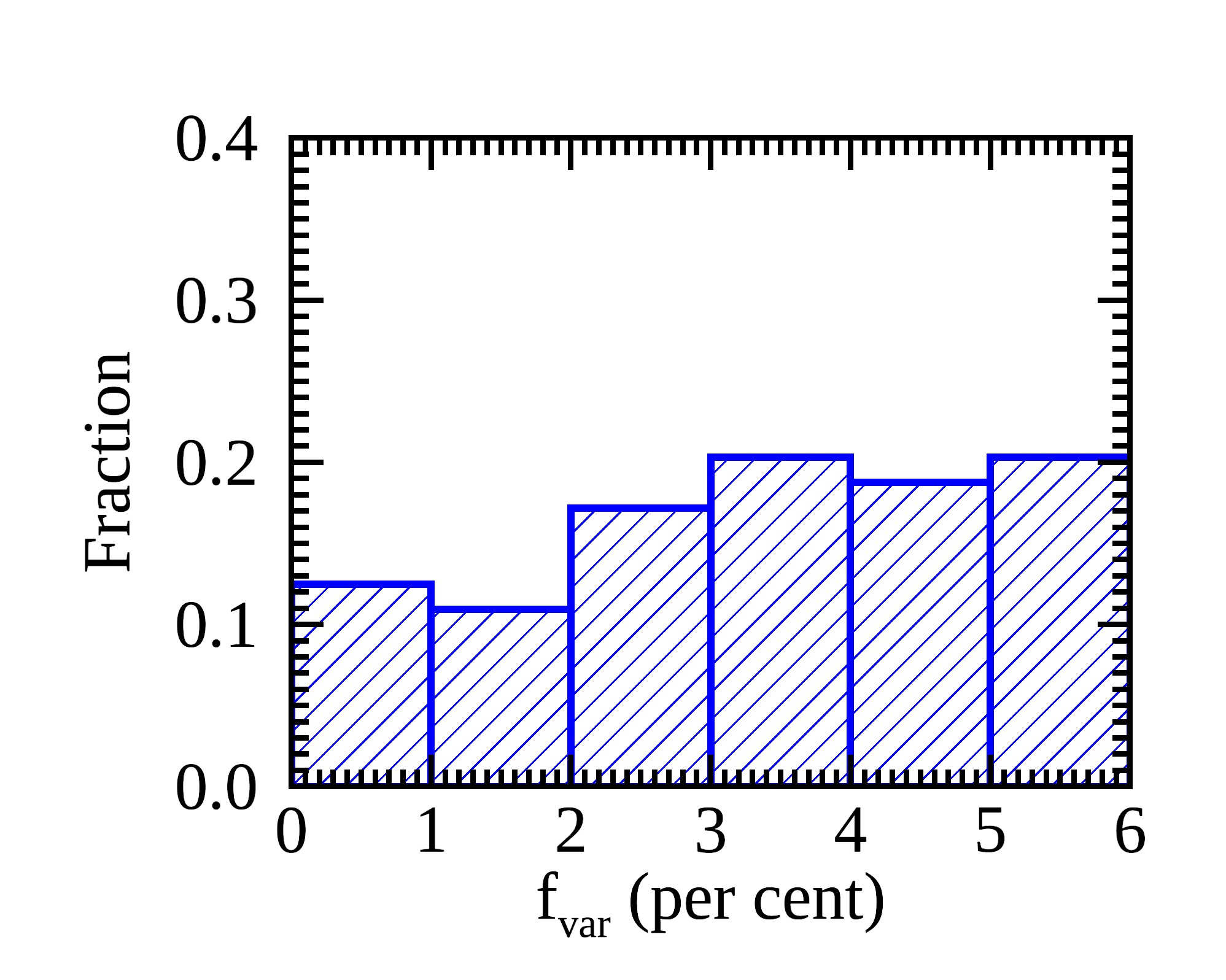,width=0.32\textwidth,clip=}
\caption{From left to right: the distributions of black hole masses, variability periods, and variability amplitudes of the mock binary sample.}
\label{fig:case}
\end{figure*} 

We then ``sample'' this binary population by considering three elements: temporal constraints, flux limit of the survey, and column density of the AGN. Since $t_{\rm res} = t_{\rm res}(t_{\rm orb}, M_{\rm BH}, q)$, we are able to calculate $t_{\rm orb}$ and therefore the observed variability timescale of each mock binary: $t_{\rm var} = t_{\rm orb}(1+z)$. Here we assume $q=0.1$, as it strikes a balance between the mass ratio expected in a major merger and the one that can cause strong periodic modulations (see below). We then only consider those with $t_{\rm var}$ that can be probed with the data length and sampling, i.e. $t_{\rm var} = [2, 730]$ days, where $t_{\rm var, max}$ is based on the assumption that at least two cycles are needed for periodicity detection and $t_{\rm var, min}$ is determined by the cadence (see Section \ref{sec:sim}). Second, we impose a soft X-ray band flux limit at $4.4\times10^{-14}$ erg s$^{-1}$cm$^{-2}$ \citep{Merloni2012}. Finally, we only consider column densities $N_{\rm H} < 10^{23}$ cm$^{-2}$, to which the soft X-ray band is sensitive. A total of $64$ binaries met these criteria. As we show in Figure \ref{fig:case}, their black hole mass distribution strongly peaks at 10$^{7-7.5} M_{\odot}$. We also find that while this sample probes the full range of input periods, longer periods between $\sim 400-600$ days are overall preferred.

We note a few caveats associated with our binary population: first, the $M_{\rm stellar}-M_{\rm BH}$ relation from \cite{Reines2015} is systematically below that of elliptical galaxies and galaxies with classical bulges, and given the strong mass dependence of the binary residence time, our final mock binary population is also dependent on our particular $M_{\rm BH}$ prescription. If an $M_{\rm stellar}-M_{\rm BH}$ relation for ellipticals is adopted instead, so that $\log(M_{\rm BH}/M_{\odot}) = 8.95+1.04 \log(M_{\rm stellar}/10^{11}M_{\odot})$ \citep{Reines2015}, the number of observable binaries decreases by a factor of $\sim 10$, due to the shorter time for a binary to evolve through the observable timescales. Additionally, since the true $M_{\rm stellar}-M_{\rm BH}$ relation for AGNs is still an active area of inquiry, the aforementioned decrease is likely only a conservative estimate. Second, we have assumed the binary evolution is primarily gravitational wave-driven, so that the residence time has a simple power-law dependence on the orbital period ($\alpha = 8/3$). However, in general, $\alpha$ is dependent on the physical mechanism driving the binary evolution, and $\alpha < 8/3$ for other processes such as gas interaction \citep{Haiman2009}. However, those mechanisms are beyond the scope of this work.


\subsection{Periodic Variability Properties}\label{sec:observe}

To calculate the expected periodic variability amplitude of each AGN, we first consider the relativistic Doppler boost model \citep{D'Orazio2015}. In this model, the line-of-sight velocity of the black hole directly translates to an apparent fractional flux variability of its minidisk emission\footnote{Here we assume the emission is dominated by the secondary black hole. See \cite{Farris2014} for an accretion prescription of individual members of the binary.}: $\Delta f/f = (3-\alpha) (v_{2}/c) \sin(i)$, where here we adopt $\alpha=1$ as the spectral index in the X-ray band (or a photon index of $\Gamma=2$), $v_{2}$ is the velocity of the secondary black hole: $v_{2} = (2\pi/1+q)(GM/4\pi^2P)^{1/3}$, and we assume random orientations of the binaries on the sky.

In the right panel of Figure \ref{fig:case}, we show the resulting distribution of the variability amplitudes. We find the periodic signals produced by this sample of binaries due to Doppler boost are at the modest level of a few percent, where only $\sim 20$\% of the SMBHBs vary at the $>5\%$ level. It can also be seen that the distribution slightly increases towards large amplitudes, but none at the $>6\%$ level are produced by this sample.

While Doppler boosting is inevitable regardless of the details of the emission, it is expected to give only a conservative estimate of the number of detectable binaries due to its strong dependence on binary parameters and the orbital inclination. Thus, we will also consider a second periodic variability model in the following section, where we assume $f_{\rm var} = 10$\% as an optimistic case. While this variability amplitude is higher than the highest amplitude of the Doppler-boosted periodic binaries in our sample, it may be more consistent with an alternative mechanism that could produce X-ray periodicity due to the outflung gas hitting the cavity wall (see \citealt{Tang2018}). Unlike Doppler boosting, the periodic amplitude in this model cannot be calculated analytically; thus, we will fix $f_{\rm var}$ at 10\% for all binary parameters and inclination angles. However, we will adopt the same $t_{\rm var}$ distribution, since it is the output of the same binary population.

\begin{figure*}
\centering
\epsfig{file=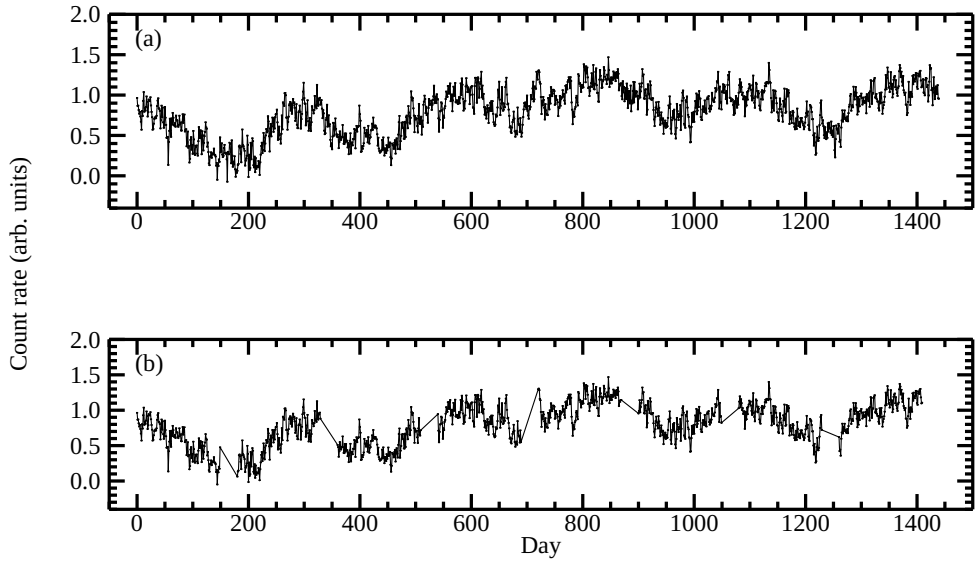,width=0.8\textwidth,clip=}
\caption{Two mock light curves down-sampled from the same parent light curve (in which a periodic signal of $P=276$ days was injected), representing two sampling cases under consideration: (a) no gap, (b) $\sim 15$\% gap. The observation times have been normalized to the first day of observations.}
\label{fig:sim_lc}
\end{figure*} 

Working under the same upper-limit assumption that binaries already exist the BAT volume, we also revisited the null detection of periodic AGNs with BAT by adopting the black hole mass and redshift measurements from BASS DR2. While BAT spans a baseline that is at least twice longer (and thus samples a wider range of $t_{\rm var}$), its measurement uncertainty is much larger than the fractional periodic variability ($\lesssim 8$\% level) of any SMBHBs, and a signal can not be detected even without the underlying red noise. This suggests that our null detection in BAT is consistent with our chosen toy model for the (upper limit) binary population.


\subsection{Light Curve Simulations}\label{sec:sim}

To investigate the detectability of a periodic signal of period $t_{\rm var}$ and amplitude $f_{\rm var}$ superimposed on red noise, we will simulate mock light curves sampled at a given cadence. We first assume an ideal, uniform cadence that the object is visited daily for the duration of the survey. However, it should be noted that this is not the actual sampling of eRASS1 -- eRASS8. During each 6-month-long full-sky scan, a given sky location will be paid several consecutive visits, which are separated by 4 hours, before eROSITA returns to it in the next eRASS. Since the current eROSITA scanning strategy is not yet publicly available, we have made a few assumptions about the sampling pattern: (1) we assume the scanning law does not evolve between eRASS1 -- eRASS8; (2) We assume that a more-densely sampled light curve can be re-binned to a daily cadence. A continuous and uniform sampling is expected to give us an upper-limit estimate of the detectability of periodic signals in red noise, and we will explore the erosion of detectability with uneven sampling in Section \ref{sec:gap}.

To produce a mock light curve of a given PSD, we use the method of \cite{Timmer1995}. Here we again adopt a single power law, since the expected PSD break $T_{\rm br}\sim$ day is shorter than the timescales probed by the daily cadence, given the black hole masses of the sample (see Section \ref{sec:analysis}). We draw the value of the PSD slope $\alpha$ from a normal distribution of $\mu = 0.9$ and $\sigma = 0.2$, which is consistent with previous studies of the PSD (e.g. \citealt{Shimizu2013}). The normalization $A$ of the single power law is such that the fractional variability is $\sim 30$\%, which is also consistent with our variable AGN sample (Section \ref{sec:analysis} and Figure \ref{fig:sigma_xs}) and largely independent of black hole mass \citep{Shimizu2013}. To mitigate possible spectral leakage, the parent light curve is $\sim 20$ times longer than the duration of the survey, i.e. $\sim 80$ yr.

\begin{table*}[ht]
\caption{Number of recovered periodic AGNs (without gaps)}
\begin{center}
\begin{tabular}{ccccccc}
\hline \hline
& \multicolumn{3}{c}{Conservative} & \multicolumn{3}{c}{Optimistic} \\
Realization & $N_{\rm tot}$  & $N_{\rm cand}$ & $N_{\rm recover}$ & $N_{\rm tot}$ & $N_{\rm cand}$ & $N_{\rm recover}$  \\
\hline
1 & 13 & 0 & 0  & 64 & 8 & 8 \\
2 & 13 & 1 & 0 & 64 & 11 & 9 \\
3 & 13 & 1 & 0 & 64 & 11 & 10 \\
4 & 13 & 0 & 0 & 64 & 7 & 6 \\
5 & 13 & 1 & 0 & 64 & 13 & 13 \\
6 & 13 & 0 & 0  & 64 & 10 & 9 \\
7 & 13 & 0 & 0 & 64 & 7 & 6 \\
8 & 13 & 1 & 0 & 64 & 9 & 8 \\
9 & 13 & 1 & 0 & 64 & 9 & 9 \\
10 & 13 & 0 & 0 & 64 & 5 & 4 \\
\hline
Average number & 13 & 0.5 & 0 & 64 & 9 & 8.2 \\
Average fraction & \nodata & 3.8\% & 0\% & \nodata & 14.1\% & 12.8\% \\
\hline \hline
\end{tabular}
\end{center}
\label{tab:gap0}
\end{table*}

Next, we inject a periodic signal, so that its period and sinusoidal amplitude are given by $t_{\rm var}$ and $f_{\rm var}$ of the mock binary, respectively. However, we only consider those with $f_{\rm var}>5$\% as our minimum signal-to-noise case, where the periodic signal has an amplitude of $f_{\rm var}$ as previously defined. Hence, the conservative Doppler model includes $13$ periodic AGNs. We have also added Poisson noise in the light curve: we assume that the measurement uncertainty between visits is negligible compared to the intrinsic stochastic variability; this corresponds to a fractional uncertainty of $\sim$ a few percent (see Section \ref{sec:analysis}). We note that this fraction is likely an underestimate for the fainter sources.

Finally, we down-sample the light curve to the cadence of each mock binary to produce the final periodic AGN mock dataset. To fully take into account the effect of red noise fluctuations, we have simulated ten realizations of this sample.

We then repeat the above light curve simulation procedure for the optimistic case, which includes $64$ AGNs. We do not require a signal-to-noise threshold in this case, since all amplitudes are fixed at 10\%. We show an example light curve in the upper panel of Figure \ref{fig:sim_lc}.


\subsection{Detectability of Periodic AGNs with Uniform Sampling}\label{sec:detect}

We then apply the method in Section \ref{sec:analysis} and search for a periodic signal at the $98$\% level, which corresponds to less than one expected false positive for our sample size. We quantify the detectability with two numbers: $N_{\rm cand}$ is the number of light curves that are identified as having peaks at this level. In a systematic search where a significance level threshold is applied, they would be selected as ``periodic candidates''. However, most of them are false positives due to red noise fluctuations, which is indicated by a high-significance peak located at the wrong frequency. Thus, if the injected periods are \emph{correctly} identified (within a 0.1 dex uncertainty from the injected value) at the $>98$\% level , we refer to the number of those as $N_{\rm recover}$.

We summarize the results in Table \ref{tab:gap0}: under the conservative periodic variability model, only one candidate can be identified in 5/10 realizations, but none of them are identified at the correct frequency. This is perhaps non-surprising, since the conservative case only contains $13$ binaries; neither does it sample a sufficient number of binaries with large amplitude periodic variations, making it challenging to identify them against red noise. In the optimistic case, where $f_{\rm var}$ is fixed at 10\% regardless of the mock binary parameter, the recovered number has significantly increased: 5-13 periodic candidates are identified in each realization of the $64$ binaries, corresponding to a candidate rate of 14.1$\pm$3.7\%. What is also noteworthy is the low false positive rate of $\sim 10$\% (represented here by $1-\frac{N_{\rm recover}}{N_{\rm cand}}$), which leads to the comparatively high recovery rate of 12.8$\pm$3.9\%. This may be due to the combination of large periodic amplitudes and the even sampling that we have adopted.

However, we stress that the recovery rates in both ``optimistic'' and ``conservative'' cases should be understood as upper limits, since (1) we have assumed that each AGN being sampled hosts an SMBHB, while the actual fraction would be much lower; (2) in both cases, we have assume that white Poisson noise is negligible with respect to red noise on the timescales of interest; (3) in our conservative case, we only consider those with large periodic amplitudes, while those with $f_{\rm var}<5$\% would likely be missed due to red noise, thus further lowering the overall recovery fraction; (4) in our optimistic case, we have fixed the periodic amplitude at $10$\%; it is therefore ``optimistic'' in the sense that the periodic amplitude is more pronounced and is independent of binary parameters.

\begin{table}[ht]
\caption{Number of recovered periodic AGNs (15\% gap)}
\begin{center}
\begin{tabular}{ccccccc}
\hline \hline
Realization & $N_{\rm tot}$  & $N_{\rm cand}$ & $N_{\rm recover}$ \\
\hline
1 & 64 & 8 & 7   \\
2 & 64 & 10 & 9  \\
3 & 64 & 9 & 8 \\
4 & 64 & 6 & 4  \\
5 & 64 & 12 & 11 \\
6 & 64 & 9 & 7  \\
7 & 64 & 9 & 8\\
8 & 64 & 8 & 7  \\
9 & 64 & 11 & 7  \\
10 & 64 & 1 & 1  \\
\hline
Average number & 64 & 8.3 & 6.9  \\
Average fraction & \nodata & 12.9\% & 10.8\%  \\
\hline \hline
\end{tabular}
\end{center}
\label{tab:gap}
\end{table}

We also note that our simple periodogram-based approach is only the preliminary step to \emph{reject} a spurious peak at a low significance level and is not meant to \emph{claim} a periodic signal at a high significance level (see \citealt{Vaughan2005}). While the former is easily applicable to a large survey dataset, in order to do the latter, better modeling of the underlying red noise and parameter uncertainties should be fully considered. Thus, while the Fourier method has the great advantage of speed and can be easily applied to a large (mock) survey dataset, our SMBHB detectability estimates presented here are not meant to replace detailed analyses involving the actual sampling.


\subsection{Mind the Gap: Detectability of Periodic AGNs with Non-uniform Sampling}\label{sec:gap}

In fact, standard Fourier methods can no longer be used for the actual sampling of eROSITA, where consecutive visits are followed by observing gaps, the length of which is a function of the latitude of the sky location. Hence, in this section, we will investigate the effects of observing gaps on both candidate and recovery rates.

To this aim, we further down-sample the same set of simulated light curves (Section \ref{sec:sim}) by inserting a 1-month-long gap every 6 months, which corresponds to $\sim 15$\% of the data being replaced with gaps over the full survey period. An example is shown in the lower panel of Figure \ref{fig:sim_lc}, which is down-sampled from the same light curve in the upper panel.

Since we can not directly apply the simple Fourier method to an unevenly-sampled light curve, we fill in the missing data by linearly interpolating across the gaps. The interpolated data are then given large ``measurement uncertainties'' which are comparable to the standard deviation of the full light curve. We assume that by replacing only 15\% of the observation length with interpolated data, any power spectrum distortion is negligible and our method in Section \ref{sec:analysis} is still valid.

\begin{figure}
\centering
\epsfig{file=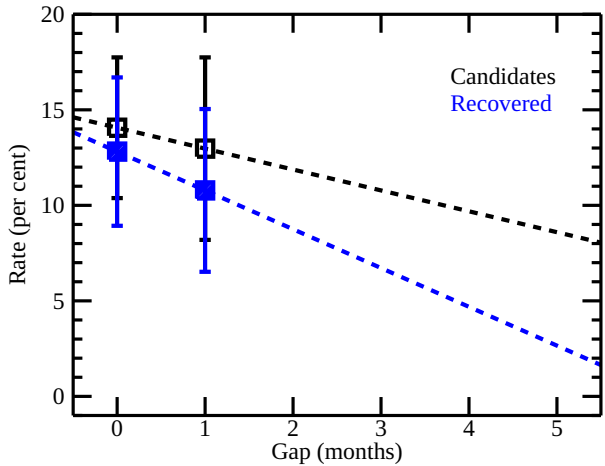,width=0.5\textwidth,clip=}
\caption{
We show the candidate and recovery rates (open black and filled blue squares, respectively) versus the length of the gap per 6-month period. The error bars represent the standard deviations of the realizations. To guide the eye, the dashed lines show the trend of decreasing rates with longer gap lengths. We note that while this trend has been linearly extrapolated to longer gaps to show the expected further decline in detectability, it does not depict the predicted candidate or recovery fraction.
}
\label{fig:rates}
\end{figure} 

We repeat the period searching procedure described Section \ref{sec:detect} for light curves in the optimistic case (i.e. 10\% periodic amplitude) and report the candidate and recovery rates in Table \ref{tab:gap}: 12.9$\pm$4.8\% of the them are identified as periodic candidates, and 10.8$\pm$4.3\% are recovered at the correct period. We then compare these detection rates with those from Section \ref{sec:detect}, where the light curves are continuously and evenly sampled. As Figure \ref{fig:rates} shows, both the number of periodic candidates and the number of recovered true periodic AGNs have decreased, which is expected due to the lack of reliable measurements during the gap periods. Further, it appears that the $N_{\rm recover}$ fraction is decreasing at a faster rate, which can be attributed to the higher number of false positives.

We hence expect that with longer gaps, the mock dataset would be less sensitive to a yet wider range of variability periods, resulting in even fewer detections. In Figure \ref{fig:rates}, for visual purposes only, we extrapolate both rates to cases with longer gaps, showing the expected further decline in the number of candidates and the number of recovered periodic sources. Unfortunately, we are unable to draw reliable conclusions about the expected detectability with long gaps, since their effects on detectability would behave in a non-linear manner for data with $> 15$\% gaps and likely render even fewer recoverable sources than the linearly-extrapolated values. We hence expect that realistic {\it eROSTIA} sampling, which is in the severely gappy regime, is unlikely to be sensitive to SMBHBs of periods of hundreds of days.


\begin{table}[ht]
\caption{Number of recovered periodic AGNs (short periods)}
\begin{center}
\begin{tabular}{ccccccc}
\hline \hline
Realization & $N_{\rm tot}$  & $N_{\rm cand}$ & $N_{\rm recover}$ \\
\hline
1 & 70 & 17 & 9 \\
2 & 70 & 20 & 15  \\
3 & 70 & 14 & 9 \\
4 & 70 & 22 & 15  \\
5 & 70 & 18 & 14 \\
6 & 70 & 17 & 15  \\
7 & 70 & 17 & 14 \\
8 & 70 & 23 & 11  \\
9 & 70 & 15 & 12 \\
10 & 70 & 24 & 16 \\
\hline
Average number & 70 & 18.7  & 13 \\
Average fraction & \nodata & 26.7\% & 18.6\% \\
\hline \hline
\end{tabular}
\end{center}
\label{tab:short}
\end{table}

\subsection{Detectability of Short Periods}\label{sec:short}

However, with 5-month gaps occurring every 6 months, we expect the sampling to be more sensitive to shorter timescales which can be probed with continuous sampling for $\sim$weeks. Hence, in this section, we will investigate the sensitivity of our sampling to short periods between 2 and 15 days. Instead of an astrophysically-motivated t$_{\rm var}$ distribution as we adopted in Section \ref{sec:mock}, we simply apply a uniform distribution, as our assumptions about the binary population only produce a statistically meaningful number of mock binaries with periods of hundreds of days. While our mock population no longer applies in the short-period case, many of the mechanisms that are expected to produce periodicities (Section \ref{sec:intro}) are still valid in this regime, since the black holes are expected to carry their mini-disks until right before merger (e.g. \citealt{Tang2018}). Since the orbital period at ISCO for a non-spinning black hole of mass 10$^{7}$ M$_{\odot}$ is $\sim$ hr, we assume that binaries of periods of days are far from the inspiral stage and that their orbits are stationary over the course of the survey.

\begin{figure}
\centering
\epsfig{file=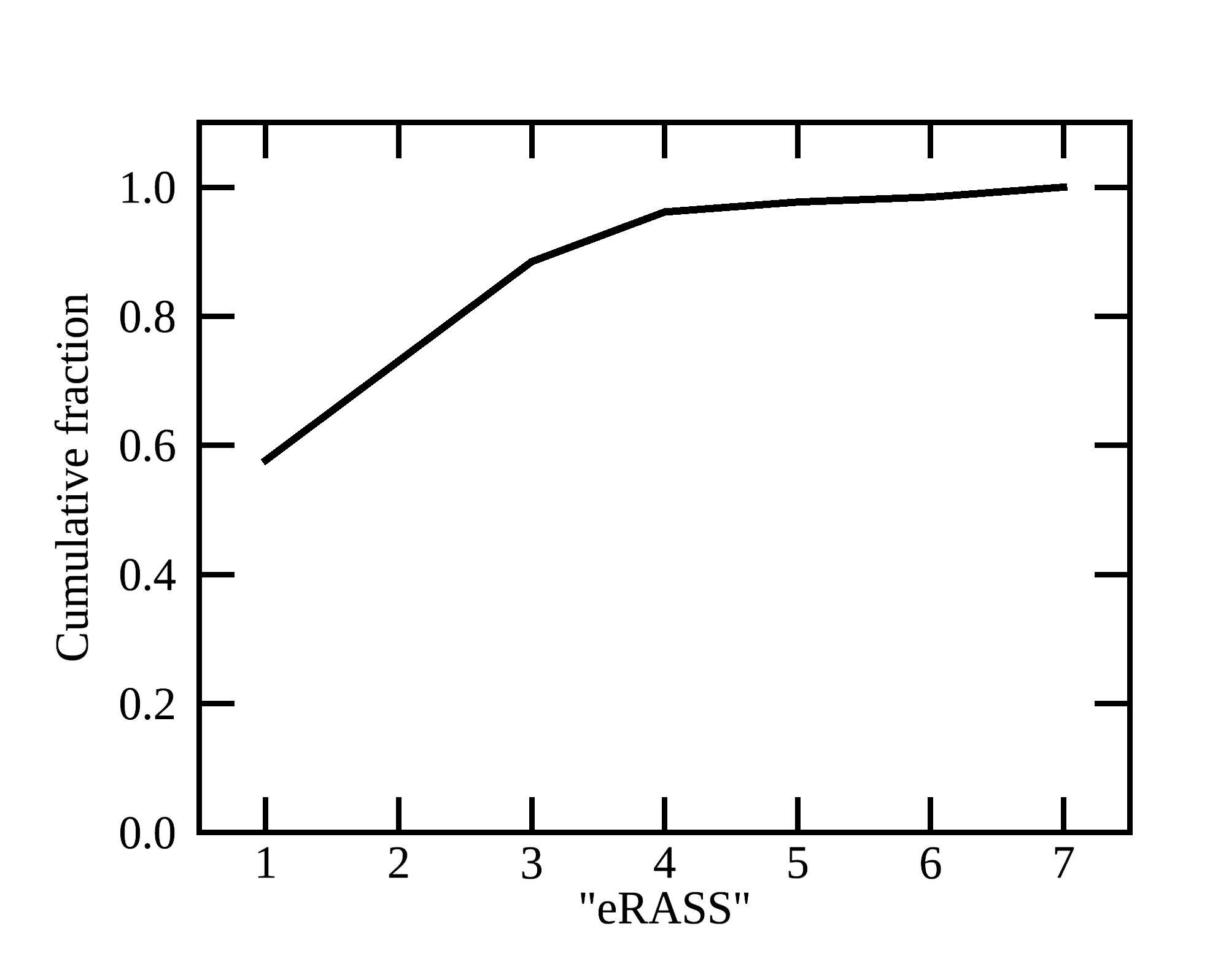,width=0.5\textwidth,clip=}
\epsfig{file=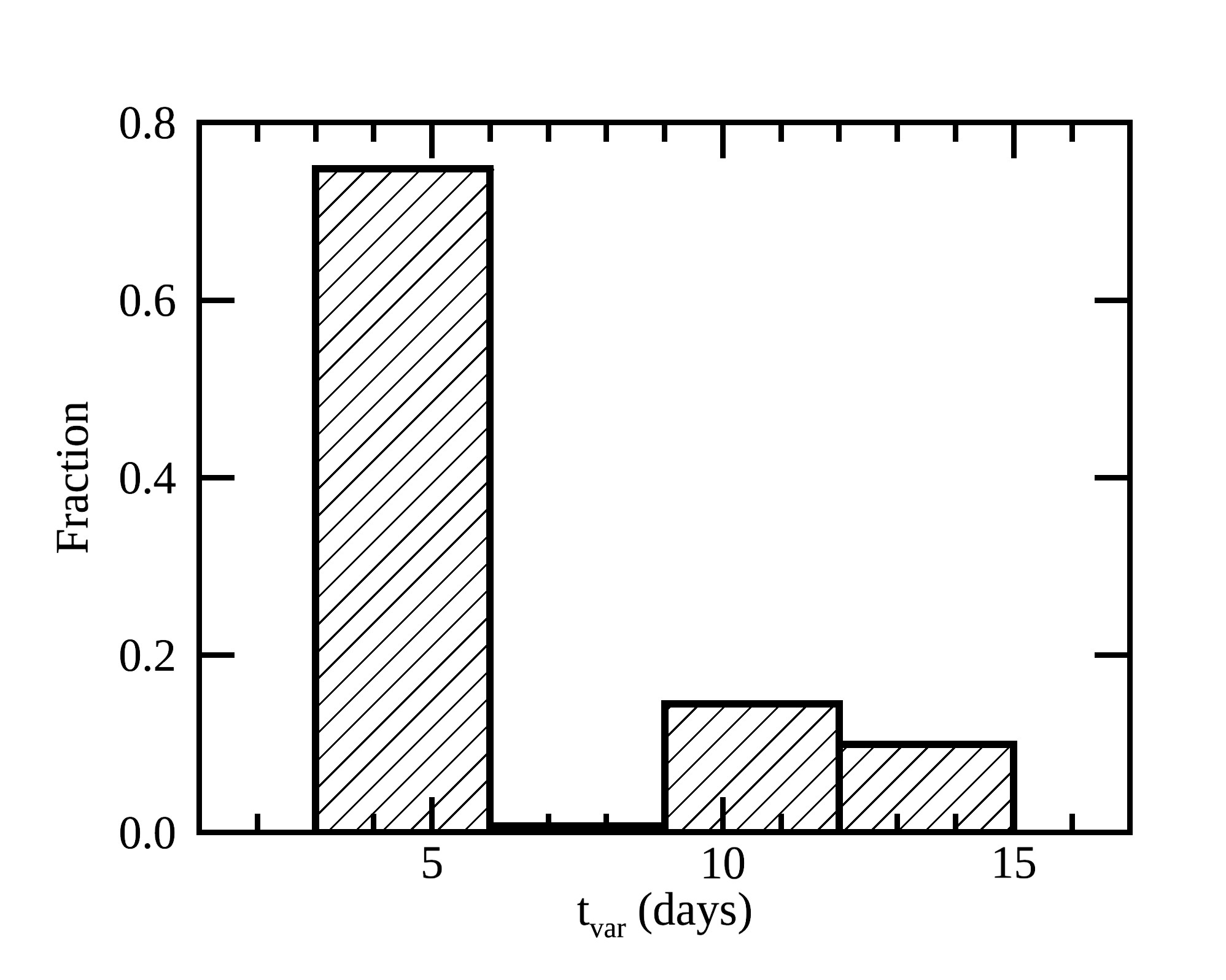,width=0.5\textwidth,clip=}
\caption{
Upper panel: the cumulative fraction of periods recovered as a function of ``eRASS''. $\sim$70\% of the signals can be recovered by the end of ``eRASS2'' at the earliest. By construction, a signal is recovered in ``eRASS7'' at the latest (see text).
Lower panel: the distribution of the recovered periods. Shorter periods are preferentially recovered due to the combination of the red noise characteristic and the total number of observed cycles. 
}
\label{fig:short}
\end{figure} 

As an optimistic estimate, we will again adopt the same fixed periodic amplitude of 10\% as in Section \ref{sec:observe}. We then followed the same procedures as we described in Section \ref{sec:sim} and generated 70 light curves for each of the ten realizations. To search for periods, we have modified our method in Section \ref{sec:detect}: we split the 4-yr-long light curve into 8 segments, each one during which the object is observed daily for one month, and each segment is treated as an independent experiment. We then stack the observations by computing the average log-periodogram after N = 1, 2 ... 8 segments, which is the analog of eRASS1--8. We further require that an object is selected as a periodic candidate or a recovered periodic source if the respective condition is met for at least two consecutive ``eRASSes''.   

We tabulate the number of candidate and recovered periods in Table \ref{tab:short}: 26.7$\pm$4.9\% of the simulations are identified as periodic candidates, and 18.6$\pm$3.7\% are recovered at the correct period, or a total number of $N_{\rm recover}  = 130$. Both rates are higher than those for long periods (Tables \ref{tab:gap0} and \ref{tab:gap}); this is likely due to the larger number of cycles for which the periodic source is observed, which improves the chance of robustly detecting the signal against red noise.

Of the 130 recovered sources, approximately 70\% are recovered by the end of ``eRASS2'', while the remaining sources are recovered as more segments are stacked (Figure \ref{fig:short}, upper panel). This is consistent with the expectation that a true periodic feature persists and its S/N gradually improves with more observations (e.g. \citealt{Liu2018}). A similar approach has also been applied in the search for QPO features to ensure the signal is stable and not due to a small number of spurious observations (e.g. \citealt{Pasham2014}). We further find that shorter periods between 3--6 days are preferentially recovered (Figure \ref{fig:short}, lower panel). In addition to the large number of cycles for which they are observed, this is also expected since the red noise level is lower at shorter timescales, making it less challenging to detect a signal of the same amplitude.


\subsection{Comparison with Previous Work}\label{sec:compare}

The electromagnetic detectability of SMBHBs has been investigated by several previous works. \cite{kelley2019} used a population of binaries from the Illustris hydrodynamical cosmological simulation and prescribed periodic variability amplitudes based on the hydrodynamical simulations by \cite{Farris2014} or Doppler boost \citep{D'Orazio2015}. They find that a current all-sky survey with a magnitude limit of $\sim 20$ mag is already capable of detecting a few binaries as hydrodynamical periodic AGNs. More encouragingly, they predict that the Large Synoptic Survey Telescope\footnote{It was announced in January 2020 that the LSST will be renamed the NSF Vera C. Rubin Observatory.} (LSST, \citealt{Ivezic2008}) can potentially discover more than a hundred periodic AGNs as SMBHBs due to either mechanism, thanks to the much larger volume that it will probe. However, the effect of the underlying red noise, which would strongly prohibit us from detecting the periodicity, has not been explored in that work.

The recent work by \cite{Krolik2019} investigated the detectability of SMBHBs either as a result of a spectral notch in the UV/optical band due to the cavity in the circumbinary disk, or an enhancement in the hard X-rays as the accretion streams shock-heat the minidisks \citep{Roedig2014}. Using a population of binaries similarly ``formed'' in a cosmological simulation, they predict that there could be $\sim 100$ binaries with X-ray flux $> 10^{-13}$ erg s$^{-1}$ cm$^{-2}$ under either model, and X-ray-enhanced binaries are a factor of a few more observable than binaries with the spectral notch feature. While the hard X-ray enhancement should also vary periodically as we discussed in Section \ref{sec:intro}, identifying such a signature is also susceptible to red noise, since the Compton reflection spectrum in the hard X-ray band should also vary in a stochastic fashion in response to the continuum.

Using an observationally-based approach, \cite{Liu2019} adopted a quasar luminosity function and an empirical variability amplitude--luminosity relation and ``selected'' variable quasars using the same pipeline as the one applied to their systematic search in the Pan-STARRS1 \citep{Kaiser2010, Chambers2016} Medium Deep Survey (PS1 MDS) and converted their upper limit from PS1 MDS to a rate for LSST: $N_{\rm LSST}<3,500$. Interestingly, if the more conservative expectation value of $N_{\rm PS1\,MDS} \sim 0.06$ based on the independent predictions by \cite{kelley2019} is adopted instead, they also arrived at a similar rate of $N_{\rm LSST} \sim 200$.

However, the effect of sampling on the periodicity detection rate was not investigated by any previous work. As we showed in Section \ref{sec:detect}, the high level of red noise and red noise fluctuations hinder the effort to identify periodic signals due to their modest amplitudes, and therefore it is quite possible that of the $100-200$ binaries predicted by \cite{kelley2019} or \cite{Krolik2019}, only a fraction can be identified observationally. The exact fraction is strongly dependent on the amplitude of the periodic signal, as we have discussed in Section \ref{sec:detect}.

The prospects for detecting SMBHBs \emph{at all} are nevertheless encouraging, given the large number of predicted SMBHBs identifiable by X-ray signatures. Assuming an optimistic recovery fraction of 13\% (Section \ref{sec:detect}) and extrapolating the same fraction to the full sky, where there are $\sim 2000$ SMBHBs with enhanced hard X-ray emissions at a few tens keV with flux $f>10^{-14}$ erg cm$^{-2}$ s$^{-1}$, $\sim 10$\% of which have short binary periods $P\lesssim5$ yr \citep{Krolik2019}, we expect to detect $\sim 26$ over ten years (we assume again that at least two cycles are required). This would require an all-sky hard X-ray survey with a sensitivity down to 10$^{-14}$ erg cm$^{-2}$ s$^{-1}$ with a few percent uncertainty; however, the detection of an enhanced hard X-ray emission, accompanied by periodicity, would provide unambiguous evidence for an SMBHB.


\section{Summary and Conclusions}\label{sec:conclude}

AGNs that host SMBHBs are predicted to vary periodically on roughly the binary orbital timescale from optical to X-rays. We have performed the first systematic search for SMBHBs in the X-rays with \emph{Swift}-BAT. While we do not find evidence for SMBHBs in the first 105 months of BAT data, including the previously reported SMBHB candidates MCG+11$-$11$-$032, we have placed an upper-limit constraint on periodic signals in the BAT volume. We further find that the lack of detections is consistent with the small expected periodic amplitudes produced by a population of SMBHBs, as well as the upper-limit detection rates inferred from previous searches in optical time-domain surveys.

We have also investigated the prospects of detecting SMBHBs with the \emph{eROSITA} survey by constructing an upper-limit population model for SMBHBs and adopting prescriptions for their periodic amplitudes. We fully take into account normal AGN X-ray variability with the red noise characteristic and investigate the detectability of those periodicities in red noise. For uniformally-sampled light curves, we find that 13\% of the periodic AGNs can be robustly identified against red noise, but the detection rate decreases with longer observing gaps. While we are unable to make solid predictions about the detectability with realistic {\it eROSTIA} sampling based on our analysis in the short-gap regime, we speculate that it is unlikely to be sensitive to bona fide SMBHBs of hundreds-of-day periods.

By contrast, short periods of days to weeks are more detectable (19\%), having benefited from more cycles being observed and the evolution of the power spectrum over time. In particular, 70\% of the recovered periods are identified by ``eRASS2'', or first year of the survey, while the remaining ones are gradually detected as the signal builds up over the course of the full survey. Shorter periods of a few days are more likely to be detected, as expected from the combination of the total number of observed cycles and the AGN red noise characteristic.


\acknowledgments

TL, MK, and GCP thank the Aspen Center for Physics, which is supported by National Science Foundation grant PHY-1607611, for hosting the ``Astrophysics of Massive Black Hole Mergers: From Galaxy Mergers to the Gravitational Wave Regime'' workshop, where many stimulating discussions took place and where this work was initiated.

TL is supported by the NANOGrav National Science Foundation Physics Frontiers Center award No. 1430284. MK acknowledges support from NASA through ADAP award NNH16CT03C. LB acknowledges support from National Science Foundation grant AST-1715413. CR acknowledges support from the CONICYT+PAI Convocatoria Nacional subvencion a instalacion en la academia convocatoria a\~{n}o 2017 PAI77170080 and Fondecyt Iniciacion grant 11190831. ET acknowledges support from CONICYT-Chile grants Basal-CATA AFB-170002, FONDECYT Regular 1160999 and 1190818, and Anillo de Ciencia y Tecnologia ACT1720033. GCP acknowledges support from the University of Florida.

This research has made use of data and/or software provided by the High Energy Astrophysics Science Archive Research Center (HEASARC), which is a service of the Astrophysics Science Division at NASA/GSFC. We acknowledge the use of public data from the Swift data archive.

\vspace{5mm}
\facilities{Swift(BAT)}



\begin{thebibliography}{}
\expandafter\ifx\csname natexlab\endcsname\relax\def\natexlab#1{#1}\fi
\providecommand{\url}[1]{\href{#1}{#1}}
\providecommand{\dodoi}[1]{doi:~\href{http://doi.org/#1}{\nolinkurl{#1}}}
\providecommand{\doeprint}[1]{\href{http://ascl.net/#1}{\nolinkurl{http://ascl.net/#1}}}
\providecommand{\doarXiv}[1]{\href{https://arxiv.org/abs/#1}{\nolinkurl{https://arxiv.org/abs/#1}}}

\bibitem[{{Ajello} {et~al.}(2010){Ajello}, {Rebusco}, {Cappelluti}, {Reimer},
  {B{\"o}hringer}, {La Parola}, \& {Cusumano}}]{Ajello2010}
{Ajello}, M., {Rebusco}, P., {Cappelluti}, N., {et~al.} 2010, \apj, 725, 1688,
  \dodoi{10.1088/0004-637X/725/2/1688}

\bibitem[{{Ajello} {et~al.}(2009){Ajello}, {Rebusco}, {Cappelluti}, {Reimer},
  {B{\"o}hringer}, {Greiner}, {Gehrels}, {Tueller}, \& {Moretti}}]{Ajello2009}
---. 2009, \apj, 690, 367, \dodoi{10.1088/0004-637X/690/1/367}

\bibitem[{{Barthelmy} {et~al.}(2005){Barthelmy}, {Barbier}, {Cummings},
  {Fenimore}, {Gehrels}, {Hullinger}, {Krimm}, {Markwardt}, {Palmer},
  {Parsons}, {Sato}, {Suzuki}, {Takahashi}, {Tashiro}, \&
  {Tueller}}]{Barthelmy2005}
{Barthelmy}, S.~D., {Barbier}, L.~M., {Cummings}, J.~R., {et~al.} 2005, \ssr,
  120, 143, \dodoi{10.1007/s11214-005-5096-3}

\bibitem[{{Baumgartner} {et~al.}(2013){Baumgartner}, {Tueller}, {Markwardt},
  {Skinner}, {Barthelmy}, {Mushotzky}, {Evans}, \& {Gehrels}}]{Baumgartner2013}
{Baumgartner}, W.~H., {Tueller}, J., {Markwardt}, C.~B., {et~al.} 2013, \apjs,
  207, 19, \dodoi{10.1088/0067-0049/207/2/19}

\bibitem[{{Beckmann} {et~al.}(2007){Beckmann}, {Barthelmy}, {Courvoisier},
  {Gehrels}, {Soldi}, {Tueller}, \& {Wendt}}]{Beckmann2007}
{Beckmann}, V., {Barthelmy}, S.~D., {Courvoisier}, T.~J.-L., {et~al.} 2007,
  \aap, 475, 827, \dodoi{10.1051/0004-6361:20078355}

\bibitem[{{Begelman} {et~al.}(1980){Begelman}, {Blandford}, \&
  {Rees}}]{Begelman1980}
{Begelman}, M.~C., {Blandford}, R.~D., \& {Rees}, M.~J. 1980, \nat, 287, 307,
  \dodoi{10.1038/287307a0}

\bibitem[{{Bird} {et~al.}(2010){Bird}, {Bazzano}, {Bassani}, {Capitanio},
  {Fiocchi}, {Hill}, {Malizia}, {McBride}, {Scaringi}, {Sguera}, {Stephen},
  {Ubertini}, {Dean}, {Lebrun}, {Terrier}, {Renaud}, {Mattana}, {G{\"o}tz},
  {Rodriguez}, {Belanger}, {Walter}, \& {Winkler}}]{Bird2010}
{Bird}, A.~J., {Bazzano}, A., {Bassani}, L., {et~al.} 2010, \apjs, 186, 1,
  \dodoi{10.1088/0067-0049/186/1/1}

\bibitem[{{Bowen} {et~al.}(2018){Bowen}, {Mewes}, {Campanelli}, {Noble},
  {Krolik}, \& {Zilh{\~a}o}}]{Bowen2018}
{Bowen}, D.~B., {Mewes}, V., {Campanelli}, M., {et~al.} 2018, \apjl, 853, L17,
  \dodoi{10.3847/2041-8213/aaa756}

\bibitem[{{Bowen} {et~al.}(2019){Bowen}, {Mewes}, {Noble}, {Avara},
  {Campanelli}, \& {Krolik}}]{Bowen2019}
{Bowen}, D.~B., {Mewes}, V., {Noble}, S.~C., {et~al.} 2019, \apj, 879, 76,
  \dodoi{10.3847/1538-4357/ab2453}

\bibitem[{{Burke-Spolaor} {et~al.}(2018){Burke-Spolaor}, {Blecha},
  {Bogdanovic}, {Comerford}, {Lazio}, {Liu}, {Maccarone}, {Pesce}, {Shen}, \&
  {Taylor}}]{Burke-Spolaor2018}
{Burke-Spolaor}, S., {Blecha}, L., {Bogdanovic}, T., {et~al.} 2018, arXiv
  e-prints.
\newblock \doarXiv{1808.04368}

\bibitem[{{Caballero-Garcia} {et~al.}(2012){Caballero-Garcia}, {Papadakis},
  {Nicastro}, \& {Ajello}}]{Caballero2012}
{Caballero-Garcia}, M.~D., {Papadakis}, I.~E., {Nicastro}, F., \& {Ajello}, M.
  2012, \aap, 537, A87, \dodoi{10.1051/0004-6361/201117974}

\bibitem[{{Chambers} {et~al.}(2016){Chambers}, {Magnier}, {Metcalfe},
  {Flewelling}, {Huber}, {Waters}, {Denneau}, {Draper}, {Farrow}, {Finkbeiner},
  {Holmberg}, {Koppenhoefer}, {Price}, {Saglia}, {Schlafly}, {Smartt},
  {Sweeney}, {Wainscoat}, {Burgett}, {Grav}, {Heasley}, {Hodapp}, {Jedicke},
  {Kaiser}, {Kudritzki}, {Luppino}, {Lupton}, {Monet}, {Morgan}, {Onaka},
  {Stubbs}, {Tonry}, {Banados}, {Bell}, {Bender}, {Bernard}, {Botticella},
  {Casertano}, {Chastel}, {Chen}, {Chen}, {Cole}, {Deacon}, {Frenk},
  {Fitzsimmons}, {Gezari}, {Goessl}, {Goggia}, {Goldman}, {Grebel}, {Hambly},
  {Hasinger}, {Heavens}, {Heckman}, {Henderson}, {Henning}, {Holman}, {Hopp},
  {Ip}, {Isani}, {Keyes}, {Koekemoer}, {Kotak}, {Long}, {Lucey}, {Liu},
  {Martin}, {McLean}, {Morganson}, {Murphy}, {Nieto-Santisteban}, {Norberg},
  {Peacock}, {Pier}, {Postman}, {Primak}, {Rae}, {Rest}, {Riess}, {Riffeser},
  {Rix}, {Roser}, {Schilbach}, {Schultz}, {Scolnic}, {Szalay}, {Seitz},
  {Shiao}, {Small}, {Smith}, {Soderblom}, {Taylor}, {Thakar}, {Thiel},
  {Thilker}, {Urata}, {Valenti}, {Walter}, {Watters}, {Werner}, {White},
  {Wood-Vasey}, \& {Wyse}}]{Chambers2016}
{Chambers}, K.~C., {Magnier}, E.~A., {Metcalfe}, N., {et~al.} 2016, ArXiv
  e-prints.
\newblock \doarXiv{1612.05560}

\bibitem[{{Charisi} {et~al.}(2016){Charisi}, {Bartos}, {Haiman},
  {Price-Whelan}, {Graham}, {Bellm}, {Laher}, \& {M{\'a}rka}}]{Charisi2016}
{Charisi}, M., {Bartos}, I., {Haiman}, Z., {et~al.} 2016, \mnras, 463, 2145,
  \dodoi{10.1093/mnras/stw1838}

\bibitem[{{Comparat} {et~al.}(2019){Comparat}, {Merloni}, {Salvato}, {Nandra},
  {Boller}, {Georgakakis}, {Finoguenov}, {Dwelly}, {Buchner}, {Del Moro},
  {Clerc}, {Wang}, {Zhao}, {Prada}, {Yepes}, {Brusa}, {Krumpe}, \&
  {Liu}}]{Comparat2019}
{Comparat}, J., {Merloni}, A., {Salvato}, M., {et~al.} 2019, \mnras, 487, 2005,
  \dodoi{10.1093/mnras/stz1390}

\bibitem[{{Di Matteo} {et~al.}(2005){Di Matteo}, {Springel}, \&
  {Hernquist}}]{DiMatteo2005Nature}
{Di Matteo}, T., {Springel}, V., \& {Hernquist}, L. 2005, \nat, 433, 604,
  \dodoi{10.1038/nature03335}

\bibitem[{{D'Orazio} {et~al.}(2013){D'Orazio}, {Haiman}, \&
  {MacFadyen}}]{D'Orazio2013}
{D'Orazio}, D.~J., {Haiman}, Z., \& {MacFadyen}, A. 2013, \mnras, 436, 2997,
  \dodoi{10.1093/mnras/stt1787}

\bibitem[{{D'Orazio} {et~al.}(2015){D'Orazio}, {Haiman}, \&
  {Schiminovich}}]{D'Orazio2015}
{D'Orazio}, D.~J., {Haiman}, Z., \& {Schiminovich}, D. 2015, \nat, 525, 351,
  \dodoi{10.1038/nature15262}

\bibitem[{{D'Orazio} \& {Loeb}(2018)}]{D'Orazio2018}
{D'Orazio}, D.~J., \& {Loeb}, A. 2018, \apj, 863, 185,
  \dodoi{10.3847/1538-4357/aad413}

\bibitem[{{Duffell} {et~al.}(2019){Duffell}, {D'Orazio}, {Derdzinski},
  {Haiman}, {MacFadyen}, {Rosen}, \& {Zrake}}]{Duffell2019}
{Duffell}, P.~C., {D'Orazio}, D., {Derdzinski}, A., {et~al.} 2019, arXiv
  e-prints, arXiv:1911.05506.
\newblock \doarXiv{1911.05506}

\bibitem[{{Edelson} {et~al.}(2002){Edelson}, {Turner}, {Pounds}, {Vaughan},
  {Markowitz}, {Marshall}, {Dobbie}, \& {Warwick}}]{Edelson2002}
{Edelson}, R., {Turner}, T.~J., {Pounds}, K., {et~al.} 2002, \apj, 568, 610,
  \dodoi{10.1086/323779}

\bibitem[{{Farris} {et~al.}(2014){Farris}, {Duffell}, {MacFadyen}, \&
  {Haiman}}]{Farris2014}
{Farris}, B.~D., {Duffell}, P., {MacFadyen}, A.~I., \& {Haiman}, Z. 2014, ApJ,
  783, 134, \dodoi{10.1088/0004-637X/783/2/134}

\bibitem[{{Farris} {et~al.}(2015){Farris}, {Duffell}, {MacFadyen}, \&
  {Haiman}}]{Farris2015}
---. 2015, \mnras, 446, L36, \dodoi{10.1093/mnrasl/slu160}

\bibitem[{{Gehrels} {et~al.}(2004){Gehrels}, {Chincarini}, {Giommi}, {Mason},
  {Nousek}, {Wells}, {White}, {Barthelmy}, {Burrows}, {Cominsky}, {Hurley},
  {Marshall}, {M{\'e}sz{\'a}ros}, {Roming}, {Angelini}, {Barbier}, {Belloni},
  {Campana}, {Caraveo}, {Chester}, {Citterio}, {Cline}, {Cropper}, {Cummings},
  {Dean}, {Feigelson}, {Fenimore}, {Frail}, {Fruchter}, {Garmire}, {Gendreau},
  {Ghisellini}, {Greiner}, {Hill}, {Hunsberger}, {Krimm}, {Kulkarni}, {Kumar},
  {Lebrun}, {Lloyd-Ronning}, {Markwardt}, {Mattson}, {Mushotzky}, {Norris},
  {Osborne}, {Paczynski}, {Palmer}, {Park}, {Parsons}, {Paul}, {Rees},
  {Reynolds}, {Rhoads}, {Sasseen}, {Schaefer}, {Short}, {Smale}, {Smith},
  {Stella}, {Tagliaferri}, {Takahashi}, {Tashiro}, {Townsley}, {Tueller},
  {Turner}, {Vietri}, {Voges}, {Ward}, {Willingale}, {Zerbi}, \&
  {Zhang}}]{Gehrels2004}
{Gehrels}, N., {Chincarini}, G., {Giommi}, P., {et~al.} 2004, \apj, 611, 1005,
  \dodoi{10.1086/422091}

\bibitem[{{Gold} {et~al.}(2014){Gold}, {Paschalidis}, {Etienne}, {Shapiro}, \&
  {Pfeiffer}}]{Gold2014}
{Gold}, R., {Paschalidis}, V., {Etienne}, Z.~B., {Shapiro}, S.~L., \&
  {Pfeiffer}, H.~P. 2014, \prd, 89, 064060, \dodoi{10.1103/PhysRevD.89.064060}

\bibitem[{{Gonz{\'a}lez-Mart{\'{\i}}n} \&
  {Vaughan}(2012)}]{Gonzalez-Martin2012}
{Gonz{\'a}lez-Mart{\'{\i}}n}, O., \& {Vaughan}, S. 2012, \aap, 544, A80,
  \dodoi{10.1051/0004-6361/201219008}

\bibitem[{{Graham} {et~al.}(2015{\natexlab{a}}){Graham}, {Djorgovski}, {Stern},
  {Glikman}, {Drake}, {Mahabal}, {Donalek}, {Larson}, \&
  {Christensen}}]{Graham2015Nat}
{Graham}, M.~J., {Djorgovski}, S.~G., {Stern}, D., {et~al.} 2015{\natexlab{a}},
  \nat, 518, 74, \dodoi{10.1038/nature14143}

\bibitem[{{Graham} {et~al.}(2015{\natexlab{b}}){Graham}, {Djorgovski}, {Stern},
  {Drake}, {Mahabal}, {Donalek}, {Glikman}, {Larson}, \&
  {Christensen}}]{Graham2015}
---. 2015{\natexlab{b}}, \mnras, 453, 1562, \dodoi{10.1093/mnras/stv1726}

\bibitem[{{Haiman}(2017)}]{Haiman2017}
{Haiman}, Z. 2017, \prd, 96, 023004, \dodoi{10.1103/PhysRevD.96.023004}

\bibitem[{{Haiman} {et~al.}(2009){Haiman}, {Kocsis}, \& {Menou}}]{Haiman2009}
{Haiman}, Z., {Kocsis}, B., \& {Menou}, K. 2009, \apj, 700, 1952,
  \dodoi{10.1088/0004-637X/700/2/1952}

\bibitem[{{Harrison} {et~al.}(2013){Harrison}, {Craig}, {Christensen},
  {Hailey}, {Zhang}, {Boggs}, {Stern}, {Cook}, {Forster}, {Giommi},
  {Grefenstette}, {Kim}, {Kitaguchi}, {Koglin}, {Madsen}, {Mao}, {Miyasaka},
  {Mori}, {Perri}, {Pivovaroff}, {Puccetti}, {Rana}, {Westergaard}, {Willis},
  {Zoglauer}, {An}, {Bachetti}, {Barri{\`e}re}, {Bellm}, {Bhalerao},
  {Brejnholt}, {Fuerst}, {Liebe}, {Markwardt}, {Nynka}, {Vogel}, {Walton},
  {Wik}, {Alexander}, {Cominsky}, {Hornschemeier}, {Hornstrup}, {Kaspi},
  {Madejski}, {Matt}, {Molendi}, {Smith}, {Tomsick}, {Ajello}, {Ballantyne},
  {Balokovi{\'c}}, {Barret}, {Bauer}, {Blandford}, {Brandt}, {Brenneman},
  {Chiang}, {Chakrabarty}, {Chenevez}, {Comastri}, {Dufour}, {Elvis}, {Fabian},
  {Farrah}, {Fryer}, {Gotthelf}, {Grindlay}, {Helfand}, {Krivonos}, {Meier},
  {Miller}, {Natalucci}, {Ogle}, {Ofek}, {Ptak}, {Reynolds}, {Rigby},
  {Tagliaferri}, {Thorsett}, {Treister}, \& {Urry}}]{Harrison2013}
{Harrison}, F.~A., {Craig}, W.~W., {Christensen}, F.~E., {et~al.} 2013, \apj,
  770, 103, \dodoi{10.1088/0004-637X/770/2/103}

\bibitem[{{Hopkins} {et~al.}(2008){Hopkins}, {Hernquist}, {Cox}, \& {Kere{\v
  s}}}]{Hopkins2008}
{Hopkins}, P.~F., {Hernquist}, L., {Cox}, T.~J., \& {Kere{\v s}}, D. 2008,
  \apjs, 175, 356, \dodoi{10.1086/524362}

\bibitem[{{Ivezic} {et~al.}(2008){Ivezic}, {Tyson}, {Abel}, {Acosta},
  {Allsman}, {AlSayyad}, {Anderson}, {Andrew}, {Angel}, {Angeli}, {Ansari},
  {Antilogus}, {Arndt}, {Astier}, {Aubourg}, {Axelrod}, {Bard}, {Barr},
  {Barrau}, {Bartlett}, {Bauman}, {Beaumont}, {Becker}, {Becla}, {Beldica},
  {Bellavia}, {Blanc}, {Blandford}, {Bloom}, {Bogart}, {Borne}, {Bosch},
  {Boutigny}, {Brandt}, {Brown}, {Bullock}, {Burchat}, {Burke}, {Cagnoli},
  {Calabrese}, {Chandrasekharan}, {Chesley}, {Cheu}, {Chiang}, {Claver},
  {Connolly}, {Cook}, {Cooray}, {Covey}, {Cribbs}, {Cui}, {Cutri}, {Daubard},
  {Daues}, {Delgado}, {Digel}, {Doherty}, {Dubois}, {Dubois-Felsmann},
  {Durech}, {Eracleous}, {Ferguson}, {Frank}, {Freemon}, {Gangler}, {Gawiser},
  {Geary}, {Gee}, {Geha}, {Gibson}, {Gilmore}, {Glanzman}, {Goodenow},
  {Gressler}, {Gris}, {Guyonnet}, {Hascall}, {Haupt}, {Hernandez}, {Hogan},
  {Huang}, {Huffer}, {Innes}, {Jacoby}, {Jain}, {Jee}, {Jernigan},
  {Jevremovic}, {Johns}, {Jones}, {Juramy-Gilles}, {Juric}, {Kahn}, {Kalirai},
  {Kallivayalil}, {Kalmbach}, {Kantor}, {Kasliwal}, {Kessler}, {Kirkby},
  {Knox}, {Kotov}, {Krabbendam}, {Krughoff}, {Kubanek}, {Kuczewski},
  {Kulkarni}, {Lambert}, {Le Guillou}, {Levine}, {Liang}, {Lim}, {Lintott},
  {Lupton}, {Mahabal}, {Marshall}, {Marshall}, {May}, {McKercher}, {Migliore},
  {Miller}, {Mills}, {Monet}, {Moniez}, {Neill}, {Nief}, {Nomerotski},
  {Nordby}, {O'Connor}, {Oliver}, {Olivier}, {Olsen}, {Ortiz}, {Owen}, {Pain},
  {Peterson}, {Petry}, {Pierfederici}, {Pietrowicz}, {Pike}, {Pinto}, {Plante},
  {Plate}, {Price}, {Prouza}, {Radeka}, {Rajagopal}, {Rasmussen}, {Regnault},
  {Ridgway}, {Ritz}, {Rosing}, {Roucelle}, {Rumore}, {Russo}, {Saha},
  {Sassolas}, {Schalk}, {Schindler}, {Schneider}, {Schumacher}, {Sebag},
  {Sembroski}, {Seppala}, {Shipsey}, {Silvestri}, {Smith}, {Smith}, {Strauss},
  {Stubbs}, {Sweeney}, {Szalay}, {Takacs}, {Thaler}, {Van Berg}, {Vanden Berk},
  {Vetter}, {Virieux}, {Xin}, {Walkowicz}, {Walter}, {Wang}, {Warner},
  {Willman}, {Wittman}, {Wolff}, {Wood-Vasey}, {Yoachim}, {Zhan}, \& {for the
  LSST Collaboration}}]{Ivezic2008}
{Ivezic}, Z., {Tyson}, J.~A., {Abel}, B., {et~al.} 2008, ArXiv e-prints.
\newblock \doarXiv{0805.2366}

\bibitem[{{Kaiser} {et~al.}(2010){Kaiser}, {Burgett}, {Chambers}, {Denneau},
  {Heasley}, {Jedicke}, {Magnier}, {Morgan}, {Onaka}, \& {Tonry}}]{Kaiser2010}
{Kaiser}, N., {Burgett}, W., {Chambers}, K., {et~al.} 2010, in Society of
  Photo-Optical Instrumentation Engineers (SPIE) Conference Series, Vol. 7733,
  Society of Photo-Optical Instrumentation Engineers (SPIE) Conference Series,
  0

\bibitem[{{Kelley} {et~al.}(2019){Kelley}, {Haiman}, {Sesana}, \&
  {Hernquist}}]{kelley2019}
{Kelley}, L.~Z., {Haiman}, Z., {Sesana}, A., \& {Hernquist}, L. 2019, \mnras,
  485, 1579, \dodoi{10.1093/mnras/stz150}

\bibitem[{{Kollmeier} {et~al.}(2017){Kollmeier}, {Zasowski}, {Rix}, {Johns},
  {Anderson}, {Drory}, {Johnson}, {Pogge}, {Bird}, {Blanc}, {Brownstein},
  {Crane}, {De Lee}, {Klaene}, {Kreckel}, {MacDonald}, {Merloni}, {Ness},
  {O'Brien}, {Sanchez-Gallego}, {Sayres}, {Shen}, {Thakar}, {Tkachenko},
  {Aerts}, {Blanton}, {Eisenstein}, {Holtzman}, {Maoz}, {Nandra}, {Rockosi},
  {Weinberg}, {Bovy}, {Casey}, {Chaname}, {Clerc}, {Conroy}, {Eracleous},
  {G{\"a}nsicke}, {Hekker}, {Horne}, {Kauffmann}, {McQuinn}, {Pellegrini},
  {Schinnerer}, {Schlafly}, {Schwope}, {Seibert}, {Teske}, \& {van
  Saders}}]{Kollmeier2017}
{Kollmeier}, J.~A., {Zasowski}, G., {Rix}, H.-W., {et~al.} 2017, arXiv
  e-prints.
\newblock \doarXiv{1711.03234}

\bibitem[{{Koss} {et~al.}(2012){Koss}, {Mushotzky}, {Treister}, {Veilleux},
  {Vasudevan}, \& {Trippe}}]{Koss2012}
{Koss}, M., {Mushotzky}, R., {Treister}, E., {et~al.} 2012, \apjl, 746, L22,
  \dodoi{10.1088/2041-8205/746/2/L22}

\bibitem[{{Koss} {et~al.}(2017){Koss}, {Trakhtenbrot}, {Ricci}, {Lamperti},
  {Oh}, {Berney}, {Schawinski}, {Balokovi{\'c}}, {Baronchelli}, {Crenshaw},
  {Fischer}, {Gehrels}, {Harrison}, {Hashimoto}, {Hogg}, {Ichikawa}, {Masetti},
  {Mushotzky}, {Sartori}, {Stern}, {Treister}, {Ueda}, {Veilleux}, \&
  {Winter}}]{Koss2017}
{Koss}, M., {Trakhtenbrot}, B., {Ricci}, C., {et~al.} 2017, \apj, 850, 74,
  \dodoi{10.3847/1538-4357/aa8ec9}

\bibitem[{{Koss} {et~al.}(2016){Koss}, {Assef}, {Balokovi{\'c}}, {Stern},
  {Gandhi}, {Lamperti}, {Alexander}, {Ballantyne}, {Bauer}, {Berney}, {Brandt},
  {Comastri}, {Gehrels}, {Harrison}, {Lansbury}, {Markwardt}, {Ricci},
  {Rivers}, {Schawinski}, {Trakhtenbrot}, {Treister}, \& {Urry}}]{Koss2016}
{Koss}, M.~J., {Assef}, R., {Balokovi{\'c}}, M., {et~al.} 2016, \apj, 825, 85,
  \dodoi{10.3847/0004-637X/825/2/85}

\bibitem[{{Koss} {et~al.}(2018){Koss}, {Blecha}, {Bernhard}, {Hung}, {Lu},
  {Trakthenbrot}, {Treister}, {Weigel}, {Sartori}, {Mushotzky}, {Schawinski},
  {Ricci}, {Veilleux}, \& {Sanders}}]{Koss2018}
{Koss}, M.~J., {Blecha}, L., {Bernhard}, P., {et~al.} 2018, \nat, 563, 214,
  \dodoi{10.1038/s41586-018-0652-7}

\bibitem[{{Krivonos} {et~al.}(2010){Krivonos}, {Tsygankov}, {Revnivtsev},
  {Grebenev}, {Churazov}, \& {Sunyaev}}]{Krivonos2010}
{Krivonos}, R., {Tsygankov}, S., {Revnivtsev}, M., {et~al.} 2010, \aap, 523,
  A61, \dodoi{10.1051/0004-6361/201014935}

\bibitem[{{Krolik} {et~al.}(2019){Krolik}, {Volonteri}, {Dubois}, \&
  {Devriendt}}]{Krolik2019}
{Krolik}, J.~H., {Volonteri}, M., {Dubois}, Y., \& {Devriendt}, J. 2019, \apj,
  879, 110, \dodoi{10.3847/1538-4357/ab24c9}

\bibitem[{{Liu} {et~al.}(2018){Liu}, {Gezari}, \& {Miller}}]{Liu2018}
{Liu}, T., {Gezari}, S., \& {Miller}, M.~C. 2018, \apjl, 859, L12,
  \dodoi{10.3847/2041-8213/aac2ed}

\bibitem[{{Liu} {et~al.}(2015){Liu}, {Gezari}, {Heinis}, {Magnier}, {Burgett},
  {Chambers}, {Flewelling}, {Huber}, {Hodapp}, {Kaiser}, {Kudritzki}, {Tonry},
  {Wainscoat}, \& {Waters}}]{Liu2015}
{Liu}, T., {Gezari}, S., {Heinis}, S., {et~al.} 2015, \apjl, 803, L16,
  \dodoi{10.1088/2041-8205/803/2/L16}

\bibitem[{{Liu} {et~al.}(2016){Liu}, {Gezari}, {Burgett}, {Chambers}, {Draper},
  {Hodapp}, {Huber}, {Kudritzki}, {Magnier}, {Metcalfe}, {Tonry}, {Wainscoat},
  \& {Waters}}]{Liu2016}
{Liu}, T., {Gezari}, S., {Burgett}, W., {et~al.} 2016, \apj, 833, 6,
  \dodoi{10.3847/0004-637X/833/1/6}

\bibitem[{{Liu} {et~al.}(2019){Liu}, {Gezari}, {Ayers}, {Burgett}, {Chambers},
  {Hodapp}, {Huber}, {Kudritzki}, {Metcalfe}, {Tonry}, {Wainscoat}, \&
  {Waters}}]{Liu2019}
{Liu}, T., {Gezari}, S., {Ayers}, M., {et~al.} 2019, \apj, 884, 36,
  \dodoi{10.3847/1538-4357/ab40cb}

\bibitem[{{MacFadyen} \& {Milosavljevi{\'c}}(2008)}]{MacFadyen2008}
{MacFadyen}, A.~I., \& {Milosavljevi{\'c}}, M. 2008, \apj, 672, 83,
  \dodoi{10.1086/523869}

\bibitem[{{Markwardt} {et~al.}(2005){Markwardt}, {Tueller}, {Skinner},
  {Gehrels}, {Barthelmy}, \& {Mushotzky}}]{Markwardt2005}
{Markwardt}, C.~B., {Tueller}, J., {Skinner}, G.~K., {et~al.} 2005, \apjl, 633,
  L77, \dodoi{10.1086/498569}

\bibitem[{{McHardy} {et~al.}(2006){McHardy}, {Koerding}, {Knigge}, {Uttley}, \&
  {Fender}}]{McHardy2006}
{McHardy}, I.~M., {Koerding}, E., {Knigge}, C., {Uttley}, P., \& {Fender},
  R.~P. 2006, \nat, 444, 730, \dodoi{10.1038/nature05389}

\bibitem[{{Merloni} {et~al.}(2012){Merloni}, {Predehl}, {Becker},
  {B{\"o}hringer}, {Boller}, {Brunner}, {Brusa}, {Dennerl}, {Freyberg},
  {Friedrich}, {Georgakakis}, {Haberl}, {Hasinger}, {Meidinger}, {Mohr},
  {Nandra}, {Rau}, {Reiprich}, {Robrade}, {Salvato}, {Santangelo}, {Sasaki},
  {Schwope}, {Wilms}, \& {German eROSITA Consortium}}]{Merloni2012}
{Merloni}, A., {Predehl}, P., {Becker}, W., {et~al.} 2012, arXiv e-prints.
\newblock \doarXiv{1209.3114}

\bibitem[{{Merloni} {et~al.}(2019){Merloni}, {Alexander}, {Banerji}, {Boller},
  {Comparat}, {Dwelly}, {Fotopoulou}, {McMahon}, {Nandra}, {Salvato}, {Croom},
  {Finoguenov}, {Krumpe}, {Lamer}, {Rosario}, {Schwope}, {Shanks}, {Steinmetz},
  {Wisotzki}, \& {Worseck}}]{Merloni2019}
{Merloni}, A., {Alexander}, D.~A., {Banerji}, M., {et~al.} 2019, arXiv
  e-prints.
\newblock \doarXiv{1903.02472}

\bibitem[{{Nandra} {et~al.}(1997){Nandra}, {George}, {Mushotzky}, {Turner}, \&
  {Yaqoob}}]{Nandra1997}
{Nandra}, K., {George}, I.~M., {Mushotzky}, R.~F., {Turner}, T.~J., \&
  {Yaqoob}, T. 1997, \apj, 476, 70, \dodoi{10.1086/303600}

\bibitem[{{Noble} {et~al.}(2012){Noble}, {Mundim}, {Nakano}, {Krolik},
  {Campanelli}, {Zlochower}, \& {Yunes}}]{Noble2012}
{Noble}, S.~C., {Mundim}, B.~C., {Nakano}, H., {et~al.} 2012, \apj, 755, 51,
  \dodoi{10.1088/0004-637X/755/1/51}

\bibitem[{{Oh} {et~al.}(2018){Oh}, {Koss}, {Markwardt}, {Schawinski},
  {Baumgartner}, {Barthelmy}, {Cenko}, {Gehrels}, {Mushotzky}, {Petulante},
  {Ricci}, {Lien}, \& {Trakhtenbrot}}]{Oh2018}
{Oh}, K., {Koss}, M., {Markwardt}, C.~B., {et~al.} 2018, \apjs, 235, 4,
  \dodoi{10.3847/1538-4365/aaa7fd}

\bibitem[{{Papadakis} \& {Lawrence}(1993)}]{Papadakis1993}
{Papadakis}, I.~E., \& {Lawrence}, A. 1993, \mnras, 261, 612,
  \dodoi{10.1093/mnras/261.3.612}

\bibitem[{{Pasham} {et~al.}(2014){Pasham}, {Strohmayer}, \&
  {Mushotzky}}]{Pasham2014}
{Pasham}, D.~R., {Strohmayer}, T.~E., \& {Mushotzky}, R.~F. 2014, \nat, 513,
  74, \dodoi{10.1038/nature13710}

\bibitem[{{Reines} \& {Volonteri}(2015)}]{Reines2015}
{Reines}, A.~E., \& {Volonteri}, M. 2015, \apj, 813, 82,
  \dodoi{10.1088/0004-637X/813/2/82}

\bibitem[{{Ricci} {et~al.}(2015){Ricci}, {Ueda}, {Koss}, {Trakhtenbrot},
  {Bauer}, \& {Gandhi}}]{Ricci2015}
{Ricci}, C., {Ueda}, Y., {Koss}, M.~J., {et~al.} 2015, \apjl, 815, L13,
  \dodoi{10.1088/2041-8205/815/1/L13}

\bibitem[{{Ricci} {et~al.}(2017{\natexlab{a}}){Ricci}, {Bauer}, {Treister},
  {Schawinski}, {Privon}, {Blecha}, {Arevalo}, {Armus}, {Harrison}, {Ho},
  {Iwasawa}, {Sanders}, \& {Stern}}]{Ricci2017}
{Ricci}, C., {Bauer}, F.~E., {Treister}, E., {et~al.} 2017{\natexlab{a}},
  \mnras, 468, 1273, \dodoi{10.1093/mnras/stx173}

\bibitem[{{Ricci} {et~al.}(2017{\natexlab{b}}){Ricci}, {Trakhtenbrot}, {Koss},
  {Ueda}, {Del Vecchio}, {Treister}, {Schawinski}, {Paltani}, {Oh}, {Lamperti},
  {Berney}, {Gandhi}, {Ichikawa}, {Bauer}, {Ho}, {Asmus}, {Beckmann}, {Soldi},
  {Balokovi{\'c}}, {Gehrels}, \& {Markwardt}}]{Ricci2017BASS}
{Ricci}, C., {Trakhtenbrot}, B., {Koss}, M.~J., {et~al.} 2017{\natexlab{b}},
  \apjs, 233, 17, \dodoi{10.3847/1538-4365/aa96ad}

\bibitem[{{Rodriguez} {et~al.}(2006){Rodriguez}, {Taylor}, {Zavala}, {Peck},
  {Pollack}, \& {Romani}}]{Rodriguez2006}
{Rodriguez}, C., {Taylor}, G.~B., {Zavala}, R.~T., {et~al.} 2006, ApJ, 646, 49,
  \dodoi{10.1086/504825}

\bibitem[{{Roedig} {et~al.}(2014){Roedig}, {Krolik}, \& {Miller}}]{Roedig2014}
{Roedig}, C., {Krolik}, J.~H., \& {Miller}, M.~C. 2014, \apj, 785, 115,
  \dodoi{10.1088/0004-637X/785/2/115}

\bibitem[{{Satyapal} {et~al.}(2017){Satyapal}, {Secrest}, {Ricci}, {Ellison},
  {Rothberg}, {Blecha}, {Constantin}, {Gliozzi}, {McNulty}, \&
  {Ferguson}}]{Satyapal2017}
{Satyapal}, S., {Secrest}, N.~J., {Ricci}, C., {et~al.} 2017, \apj, 848, 126,
  \dodoi{10.3847/1538-4357/aa88ca}

\bibitem[{{Segreto} {et~al.}(2010){Segreto}, {Cusumano}, {Ferrigno}, {La
  Parola}, {Mangano}, {Mineo}, \& {Romano}}]{Segreto2010}
{Segreto}, A., {Cusumano}, G., {Ferrigno}, C., {et~al.} 2010, \aap, 510, A47,
  \dodoi{10.1051/0004-6361/200911779}

\bibitem[{{Severgnini} {et~al.}(2018){Severgnini}, {Cicone}, {Della Ceca},
  {Braito}, {Caccianiga}, {Ballo}, {Campana}, {Moretti}, {La Parola},
  {Vignali}, {Zaino}, {Matzeu}, \& {Landoni}}]{Severgnini2018}
{Severgnini}, P., {Cicone}, C., {Della Ceca}, R., {et~al.} 2018, \mnras, 479,
  3804, \dodoi{10.1093/mnras/sty1699}

\bibitem[{{Shi} \& {Krolik}(2016)}]{Shi2016}
{Shi}, J.-M., \& {Krolik}, J.~H. 2016, \apj, 832, 22,
  \dodoi{10.3847/0004-637X/832/1/22}

\bibitem[{{Shi} {et~al.}(2012){Shi}, {Krolik}, {Lubow}, \& {Hawley}}]{Shi2012}
{Shi}, J.-M., {Krolik}, J.~H., {Lubow}, S.~H., \& {Hawley}, J.~F. 2012, \apj,
  749, 118, \dodoi{10.1088/0004-637X/749/2/118}

\bibitem[{{Shimizu} \& {Mushotzky}(2013)}]{Shimizu2013}
{Shimizu}, T.~T., \& {Mushotzky}, R.~F. 2013, \apj, 770, 60,
  \dodoi{10.1088/0004-637X/770/1/60}

\bibitem[{{Soldi} {et~al.}(2014){Soldi}, {Beckmann}, {Baumgartner}, {Ponti},
  {Shrader}, {Lubi{\'n}ski}, {Krimm}, {Mattana}, \& {Tueller}}]{Soldi2014}
{Soldi}, S., {Beckmann}, V., {Baumgartner}, W.~H., {et~al.} 2014, \aap, 563,
  A57, \dodoi{10.1051/0004-6361/201322653}

\bibitem[{{Tang} {et~al.}(2018){Tang}, {Haiman}, \& {MacFadyen}}]{Tang2018}
{Tang}, Y., {Haiman}, Z., \& {MacFadyen}, A. 2018, \mnras, 476, 2249,
  \dodoi{10.1093/mnras/sty423}

\bibitem[{{Timmer} \& {Koenig}(1995)}]{Timmer1995}
{Timmer}, J., \& {Koenig}, M. 1995, \aap, 300, 707

\bibitem[{{Tueller} {et~al.}(2008){Tueller}, {Mushotzky}, {Barthelmy},
  {Cannizzo}, {Gehrels}, {Markwardt}, {Skinner}, \& {Winter}}]{Tueller2008}
{Tueller}, J., {Mushotzky}, R.~F., {Barthelmy}, S., {et~al.} 2008, \apj, 681,
  113, \dodoi{10.1086/588458}

\bibitem[{{Tueller} {et~al.}(2010){Tueller}, {Baumgartner}, {Markwardt},
  {Skinner}, {Mushotzky}, {Ajello}, {Barthelmy}, {Beardmore}, {Brandt},
  {Burrows}, {Chincarini}, {Campana}, {Cummings}, {Cusumano}, {Evans},
  {Fenimore}, {Gehrels}, {Godet}, {Grupe}, {Holland}, {Kennea}, {Krimm},
  {Koss}, {Moretti}, {Mukai}, {Osborne}, {Okajima}, {Pagani}, {Page}, {Palmer},
  {Parsons}, {Schneider}, {Sakamoto}, {Sambruna}, {Sato}, {Stamatikos},
  {Stroh}, {Ukwata}, \& {Winter}}]{Tueller2010}
{Tueller}, J., {Baumgartner}, W.~H., {Markwardt}, C.~B., {et~al.} 2010, \apjs,
  186, 378, \dodoi{10.1088/0067-0049/186/2/378}

\bibitem[{{Ubertini} {et~al.}(2003){Ubertini}, {Lebrun}, {Di Cocco}, {Bazzano},
  {Bird}, {Broenstad}, {Goldwurm}, {La Rosa}, {Labanti}, {Laurent}, {Mirabel},
  {Quadrini}, {Ramsey}, {Reglero}, {Sabau}, {Sacco}, {Staubert}, {Vigroux},
  {Weisskopf}, \& {Zdziarski}}]{Ubertini2003}
{Ubertini}, P., {Lebrun}, F., {Di Cocco}, G., {et~al.} 2003, \aap, 411, L131,
  \dodoi{10.1051/0004-6361:20031224}

\bibitem[{{Vaughan}(2005)}]{Vaughan2005}
{Vaughan}, S. 2005, \aap, 431, 391, \dodoi{10.1051/0004-6361:20041453}

\bibitem[{{Vaughan} {et~al.}(2003){Vaughan}, {Edelson}, {Warwick}, \&
  {Uttley}}]{Vaughan2003}
{Vaughan}, S., {Edelson}, R., {Warwick}, R.~S., \& {Uttley}, P. 2003, \mnras,
  345, 1271, \dodoi{10.1046/j.1365-2966.2003.07042.x}

\bibitem[{{Vaughan} {et~al.}(2016){Vaughan}, {Uttley}, {Markowitz},
  {Huppenkothen}, {Middleton}, {Alston}, {Scargle}, \& {Farr}}]{Vaughan2016}
{Vaughan}, S., {Uttley}, P., {Markowitz}, A.~G., {et~al.} 2016, \mnras, 461,
  3145, \dodoi{10.1093/mnras/stw1412}

\bibitem[{{Wik} {et~al.}(2011){Wik}, {Sarazin}, {Finoguenov}, {Baumgartner},
  {Mushotzky}, {Okajima}, {Tueller}, \& {Clarke}}]{Wik2011}
{Wik}, D.~R., {Sarazin}, C.~L., {Finoguenov}, A., {et~al.} 2011, \apj, 727,
  119, \dodoi{10.1088/0004-637X/727/2/119}

\bibitem[{{Winkler} {et~al.}(2003){Winkler}, {Courvoisier}, {Di Cocco},
  {Gehrels}, {Gim{\'e}nez}, {Grebenev}, {Hermsen}, {Mas-Hesse}, {Lebrun},
  {Lund}, {Palumbo}, {Paul}, {Roques}, {Schnopper}, {Sch{\"o}nfelder},
  {Sunyaev}, {Teegarden}, {Ubertini}, {Vedrenne}, \& {Dean}}]{Winkler2003}
{Winkler}, C., {Courvoisier}, T.~J.-L., {Di Cocco}, G., {et~al.} 2003, \aap,
  411, L1, \dodoi{10.1051/0004-6361:20031288}

\end{thebibliography}


\end{document}